\documentclass[preprint]{aastex}

\usepackage{etex}   

\usepackage[english]{babel}
\usepackage{graphicx}
\usepackage{subfigure}
\usepackage{placeins}  

\usepackage{array}   
\usepackage{booktabs} 
\usepackage{multirow} 
\usepackage{mathtools}

\usepackage{color}  
\usepackage[percent]{overpic}

\usepackage{multicol}
\usepackage{adjustbox} 

\usepackage{natbib}

\usepackage{tikz}    
\usepackage{cancel}
\usepackage[normalem]{ulem}  

\usepackage{hyperref}    
\usepackage{cleveref} 

\usepackage{color, colortbl}
\definecolor{Gray}{gray}{0.9}
\definecolor{LightCyan}{rgb}{0.88,1,1}

\usetikzlibrary{matrix}

\shorttitle{X-raying the dark side of Venus}
\shortauthors{Afshari et al.}

\begin{document}


\title{X-Raying the Dark Side of Venus - Scatter from Venus Magnetotail?}


\author{M. Afshari\altaffilmark{1,2}, G. Peres\altaffilmark{1,2}, P. R. Jibben\altaffilmark{3}, A. Petralia\altaffilmark{1,2}, F. Reale\altaffilmark{1,2}, M. Weber\altaffilmark{3}}
\affil{\altaffilmark{1}Dipartimento di Fisica e Chimica, Universit\`a di Palermo, Piazza del Parlamento 1, 90134, Italy 
\email{peres@astropa.unipa.it}}


\affil{\altaffilmark{2} INAF- Osservatorio Astronomico di Palermo, Palermo, Piazza del Parlamento 1, 90134, Italy}

\affil{\altaffilmark{3}Harvard-Smithsonian Center for Astrophysics, 60 Garden Street, Cambridge, MA 02138, USA}


\begin{abstract}

This work analyzes the X-ray, EUV and UV emission apparently coming from the
Earth-facing (dark) side of Venus as observed with \textit{Hinode}/XRT
and SDO/AIA during a transit across the solar disk occurred in
2012. We have measured significant
X-Ray, EUV and UV flux from Venus\rq\ dark side. As a check we have
also analyzed a Mercury transit across the solar disk, observed with
\textit{Hinode}/XRT in 2006.  We have used the latest version of the
\textit{Hinode}/XRT Point Spread Function (PSF) to deconvolve Venus
and Mercury X-ray images, in order to remove possible instrumental
scattering.  Even after deconvolution, the flux from Venus\rq\  shadow
remains significant while in the case of Mercury it becomes negligible.
Since stray-light contamination affects the XRT Ti-poly filter data
from the Venus transit in 2012, we performed the same analysis with
XRT Al-mesh filter data, which is not affected by the light leak. Even
the Al-mesh filter data show residual flux.\\
We have also found significant EUV (304 \AA, 193 \AA, 335 \AA) and UV
(1700 \AA) flux in Venus\rq\ shadow, as measured with SDO/AIA.
The EUV emission from Venus\rq\ dark side is reduced when appropriate
deconvolution methods are applied; the emission remains significant,
however.\\
The light curves of the average flux of the shadow in the X-ray, 
EUV, and UV bands appear different as Venus crosses the solar disk,
but in any of them the flux is, at any time, approximately proportional to
the average flux in a ring surrounding Venus, and therefore proportional
to the average flux of the solar regions around Venus\rq\ obscuring
disk line of sight. The proportionality factor depends on the band.\\
This phenomenon has no clear origin; we suggest it may be due to scatter
occurring in the very long magnetotail of Venus. 
\end{abstract}

\keywords{\small Venus; Mercury; X-rays; Deconvolution; Hinode/XRT;
SDO/AIA; magnetotail}

\section{Introduction}
Transits of Mercury and Venus across the solar disk are well-observed celestial
phenomena. Recently, the
transit of Mercury observed with \textit{Hinode}/X-Ray Telescope (XRT;
\citealt{Gol07}) has been used by \citet{Web07} to test the sharpness
of the instrument Point Spread Function (PSF). \citet{rea15} used
\textit{Hinode}/XRT observations of a Venus transit to measure the size of
Venus in the X-ray band thus inferring the extension and optical thickness
of Venus\rq\ atmosphere. The methods and implications of the latter work reach
into planetary physics and hint at similar methods to be potentially used,
in the future, for exoplanets.\\
In this
work we analyze the same set of observations to explore the residual
X-ray emission observed in Venus\rq\ shadow and find, with the help of
an updated version of the \textit{Hinode}/XRT PSF, that this emission is not
due to instrumental scattering and may have an origin more directly related
to Venus. Previous observations with Chandra
in 2001 and then in 2006/2007 confirmed the X-ray emission from
the sunlit side of the Venus (\citealt{Den02} and  \citealt{den08}).\\
In Section 2 we present the observations of Mercury and Venus with a
brief summary of the satellites and their instruments; in Section 3 we
measure the residual flux in the shadow of Mercury in X-ray and of Venus
in X-Ray, EUV and UV bands, and its evolution as Venus crosses the solar disk. In Section
4 we deconvolve X-ray images using the updated PSF and different codes, and
again explore similarities and differences among the various observations;
in Section 5 we describe the XRT straylight contamination and present our results
taken with the Al-mesh filter. In Section 6 we show similar results
obtained in EUV and UV bands. Section 7 contains our discussion and
the conclusions.

\section{Observation: Transit of Mercury and Venus}
\begin{figure} [!t]            

 \begin{minipage}[t]{0.4\linewidth}
 \includegraphics[trim={5cm 3cm 3cm 1cm},clip,width=\textwidth]{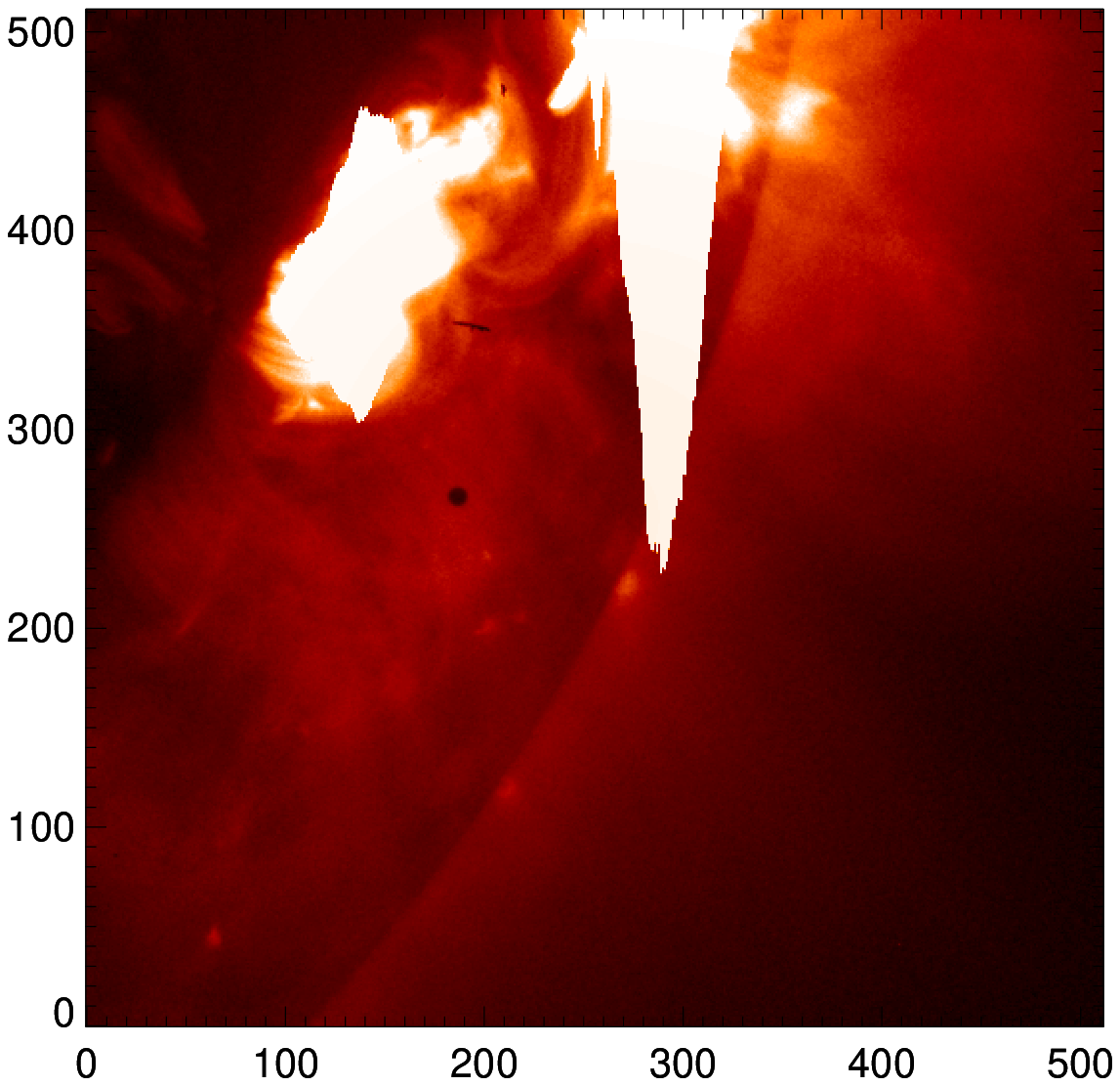}
 \hspace{0.1cm}
  \end{minipage}%
 \hspace{.05\linewidth}
\begin{minipage}[t]{0.37\linewidth}
 \includegraphics[trim={5cm 2cm 5.5cm 3.5cm},clip,width=\textwidth]{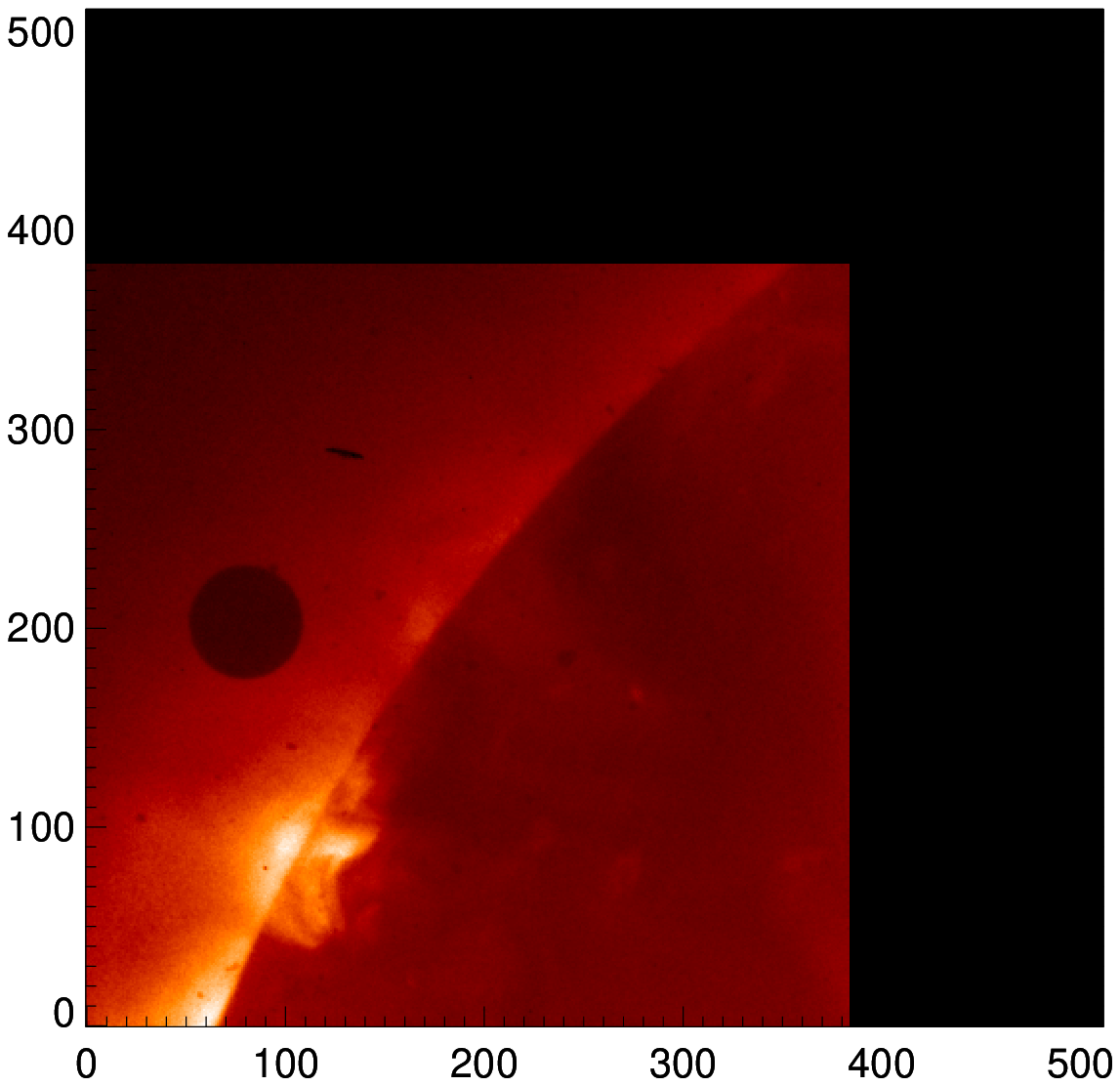}
 \hspace{0.1cm}
  \end{minipage}%

 \caption{\small Left: Mercury transit across the Sun observed with \textit{Hinode}/XRT in the X-Ray band. \newline (Time of observation, 2006-11-08 23:51:04.571).\newline
Right: Venus (black circle) approaching the Sun, observed with \textit{Hinode}/XRT in the X-Ray band. \newline
(Time of observation 2012-06-05 21:57:39.893). }
\label{transit}
\end{figure}

On 2006 Nov 08,
Mercury passed across the solar disk. Its transit lasted for almost
five hours and was observed with \textit{Hinode}/XRT in the X-ray band;
Fig.~\ref{transit} shows a selected image of this phenomenon.\\
A Venus transit was observed with \textit{Hinode}/XRT
in 2012 while it was crossing the northern hemisphere of the Sun; the
transit lasted over six hours. On the 5th of June 2012, the Venus transit began
at 22:09 UTC and finished on June 6th at 04:49 UTC. The Venus 
transit was also observed with the \textit{Solar Dynamics
Observatory}/Atmospheric Imaging Assembly (SDO/AIA) \citep{Pes12}
in the Ultraviolet (UV) and Extreme Ultraviolet (EUV) bands. Fig.~\ref{transit}
shows an image taken during this transit.\\
In the following we briefly discuss the
satellites and the instruments which took the data used in this work.\\

\subsection{Hinode/XRT}
The \textit{Hinode} satellite (formerly Solar-B) of the Japan Aerospace
Exploration Agency's Institute of Space and Astronautical Science
(ISAS/JAXA) was successfully launched in September 2006. There are three
instruments onboard: the Solar Optical Telescope (SOT), the EUV Imaging
Spectrometer (EIS), and the X-Ray Telescope (XRT). We used only data
from XRT.\\ XRT is a high-resolution grazing-incidence telescope with a
modified Wolter-I telescope design that uses grazing incidence optics
with an angular resolution consistent with 1.0286 arcsec per pixel at
the CCD \citep{Gol07}. \\ An improved version of the \textit{Hinode}/XRT PSF has been derived by
P. R. Jibben of the XRT instrument team. The Model PSF has 99\% of the encircled energy within a
100 arcsec diameter with the remaining 1\% scattered beyond\footnote{For
more information about the derivation of the PSF, the interested reader
can refer to Appendix A.}. The PSF model at 0.56 keV is:

\begin{doublespace} 
\[
    PSF= 
\begin{cases}
    
    a \frac{exp(-\frac{r^2}{\sigma^2})}{\gamma^2+r^2},& \text{if $ r  \leq 3.4176 $}; \\
    \frac{0.03}{r},& \text{if $3.4176 \leq r  \leq 5 $};  \\
    \frac{0.15}{r^2},& \text{if $5 \leq r  \leq 11.1 $}; \\
    \frac{(11.1)^2\times0.15}{r^4},& \text{if $ r  \geq 11.1 $}; 
   
\end{cases}
\]

\normalsize   
\end{doublespace}

Where r = radial distance in arc seconds, a = 1.31946, \(\sigma = 2.19256 \) and \(\gamma= 1.24891\).\\

This PSF is planned for distribution in the XRT branch of SolarSoft \citep{Fre00} \citep{Ben98}.\\

\subsection{SDO/AIA}
The \textit{Solar Dynamics Observatory} (SDO)  was launched on February
11, 2010. The spacecraft includes three instruments: the Extreme Ultraviolet
Variability Experiment (EVE), the Helioseismic and Magnetic Imager (HMI),
and the Atmospheric Imaging Assembly (AIA) \citep{Lem12}. We used only data
taken with AIA.\\ AIA, with an angular resolution of 0.6 arcsec per pixel,
provides narrow-band imaging in seven extreme ultraviolet (EUV) band
passes centered on specific lines: (94  {\AA}, 131  {\AA}, 171  {\AA},
193  {\AA}, 211  {\AA}, 304  {\AA} and 335  {\AA}) and in two UV
band-passes near 1600  {\AA} and 1700  {\AA} \citep{Lem12}.

\subsection{Data sets}
For Venus\rq\  shadow analysis, we used six different data sets in the X-ray
band, each with more than 300 images, and four data sets from AIA: at
1700 {\AA}, 335 {\AA}, 304 {\AA}, 
193 {\AA}, respectively with 114, 169, 118 and 119 images. For the Mercury
shadow analysis we used one
data set in the X-ray band. A summary of the data sets is presented in
Table 1.
The filters for all selected
images of Venus in the X-ray band are Ti-poly and Al-Mesh, and Al-poly
for Mercury images. (Ti-poly and Al-poly are metal foils on a polyimide
substrate, and Al-mesh is an Al foil mounted on a fine stainless steel mesh.)
The field of view is $384 \times 384$ pixels for Ti-poly and Al-poly images, and
is $192 \times 192$ pixels for the Al-mesh images (where each CCD pixel has been
summed $2 \times 2$). The AIA and
XRT plate scales are 0.6 arcsec per pixel and 1.0286 arcsec per pixel (.0572 arcsec per pixel for Al-mesh),
respectively.\\ \textit{Hinode}/XRT didn\rq t take any full solar
disk images of the Venus transit but only partial images of the disk where Venus
was. For the data analysis we used the standard instrumental calibration
routines provided through SolarSoft.
\begin{table}[htp!]    
\caption {Summary of data sets of Venus and Mercury} 
          
\begin{tabular}{ |  >{\centering}m{3.5em} |  >{\centering}m{1.8cm}|  >{\centering}m{2.3cm} |   >{\centering}m{5cm}|  >{\centering}m{5cm} | } 
\hline 
Planet & Filter & Instrument & Start Time of observation (UTC Time) & 
Final Time of observation (UTC  Time) \tabularnewline 
\hline
Venus & Ti-poly & \textit{Hinode}/XRT & 2012-06-05T20:03:00.615&
2012-06-05T21:58:33.335\tabularnewline 
\hline
Venus & Ti-poly & \textit{Hinode}/XRT &2012-06-05T21:58:39.912&
2012-06-06T00:23:37.912\tabularnewline 
\hline
Venus & Ti-poly & \textit{Hinode}/XRT &2012-06-06T00:23:57.272&
2012-06-06T02:06:39.223\tabularnewline
\hline
Venus & Ti-poly & \textit{Hinode}/XRT &2012-06-06T02:06:57.299&
2012-06-06T03:51:08.500 \tabularnewline 
\hline
Venus & Ti-poly & \textit{Hinode}/XRT &2012-06-06T03:51:27.859&
2012-06-06T06:47:15.490\tabularnewline
\hline
Venus & 193{\AA}&SDO/AIA &2012-06-05T22:23:07.84&
 2012-06-06T04:17:07.84 \tabularnewline
\hline
Venus & 304{\AA}&SDO/AIA &2012-06-05T22:23:08.13&
 2012-06-06T04:17:08.12\tabularnewline
 \hline
Venus & 335{\AA}&SDO/AIA &2012-06-05T22:25:03.62&
 2012-06-06T04:01:03.62 \tabularnewline
 \hline
Venus & 1700{\AA}&SDO/AIA &2012-06-05T22:32:07.71&
 2012-06-06T04:11:19.71 \tabularnewline
 \hline
Venus & Al-mesh & \textit{Hinode}/XRT & 2012-06-05T21:06:28.326&
2012-06-06T06:44:46.712\tabularnewline
 \hline
Mercury & Al-poly & \textit{Hinode}/XRT & 2006-11-08T23:50:12.052&
2006-11-08T23:59:16.234.\tabularnewline \hline

\end{tabular}

\normalsize   

\end{table}
\section{Data Analysis}
To analyze the features of Venus\rq\ and Mercury\rq s shadows in the X-Ray
band we have measured, in each image, the flux across the planetary disk and
in the nearby solar disk regions. To illustrate the features of such an
emission we show the average flux measured along strips 3 pixels wide
(in order to have a significant S/N ratio). We have considered strips
along the planet\rq s diameters, along both the N-S (vertical) and the E-W
(horizontal) directions.

\subsection{Venus Intensity Profile Analysis} 
\begin{figure}[!h]       
\centering
\begin{minipage}[]{0.25\linewidth}
 \includegraphics[trim={5cm 2cm 5.5cm 3.5cm},clip,width=\textwidth]{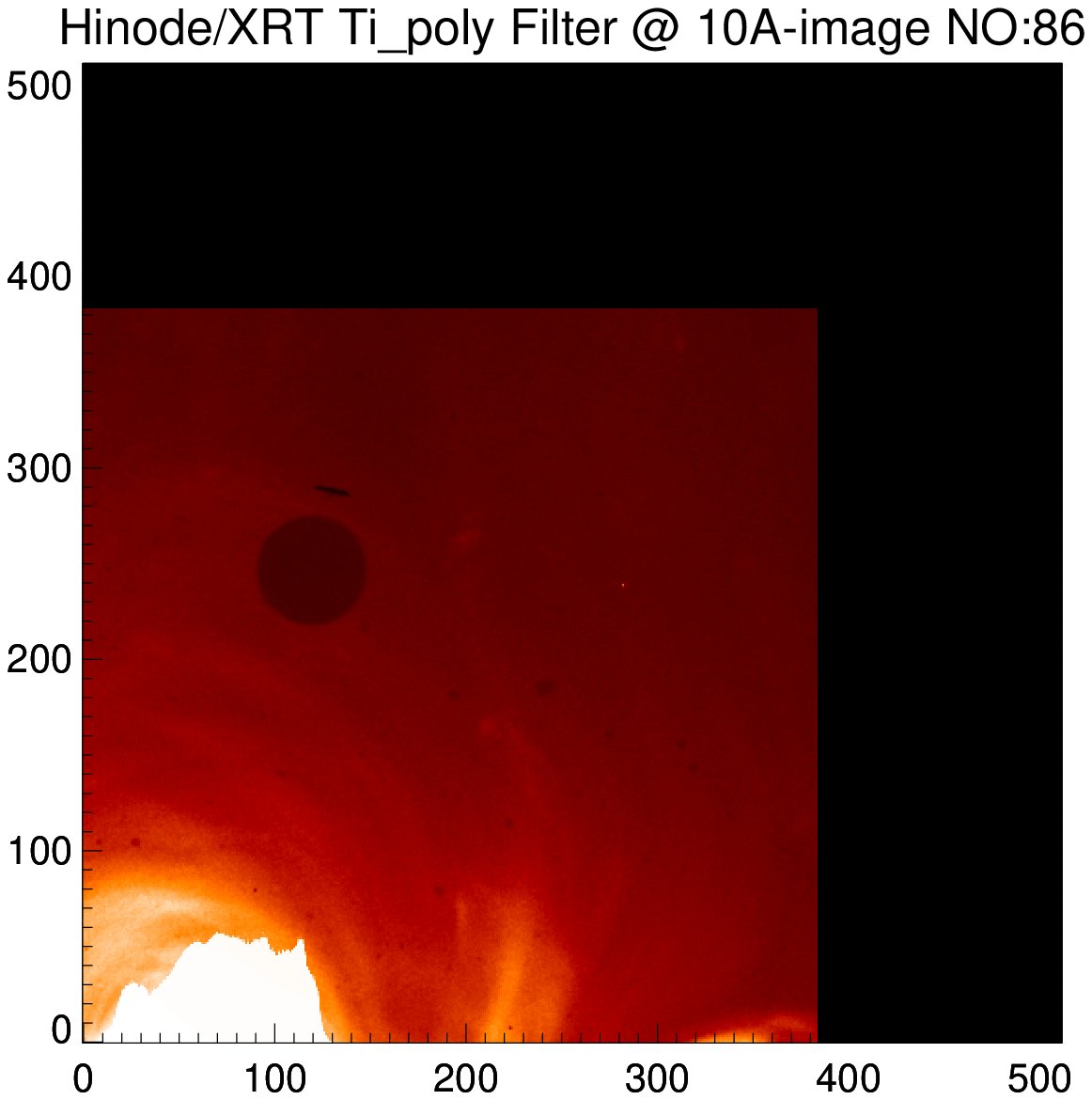}
   \hspace{10cm}   
  \end{minipage}
\begin{minipage}[]{0.25\linewidth}
 \includegraphics[trim={5cm 2cm 4cm 2cm},clip,width=\textwidth]{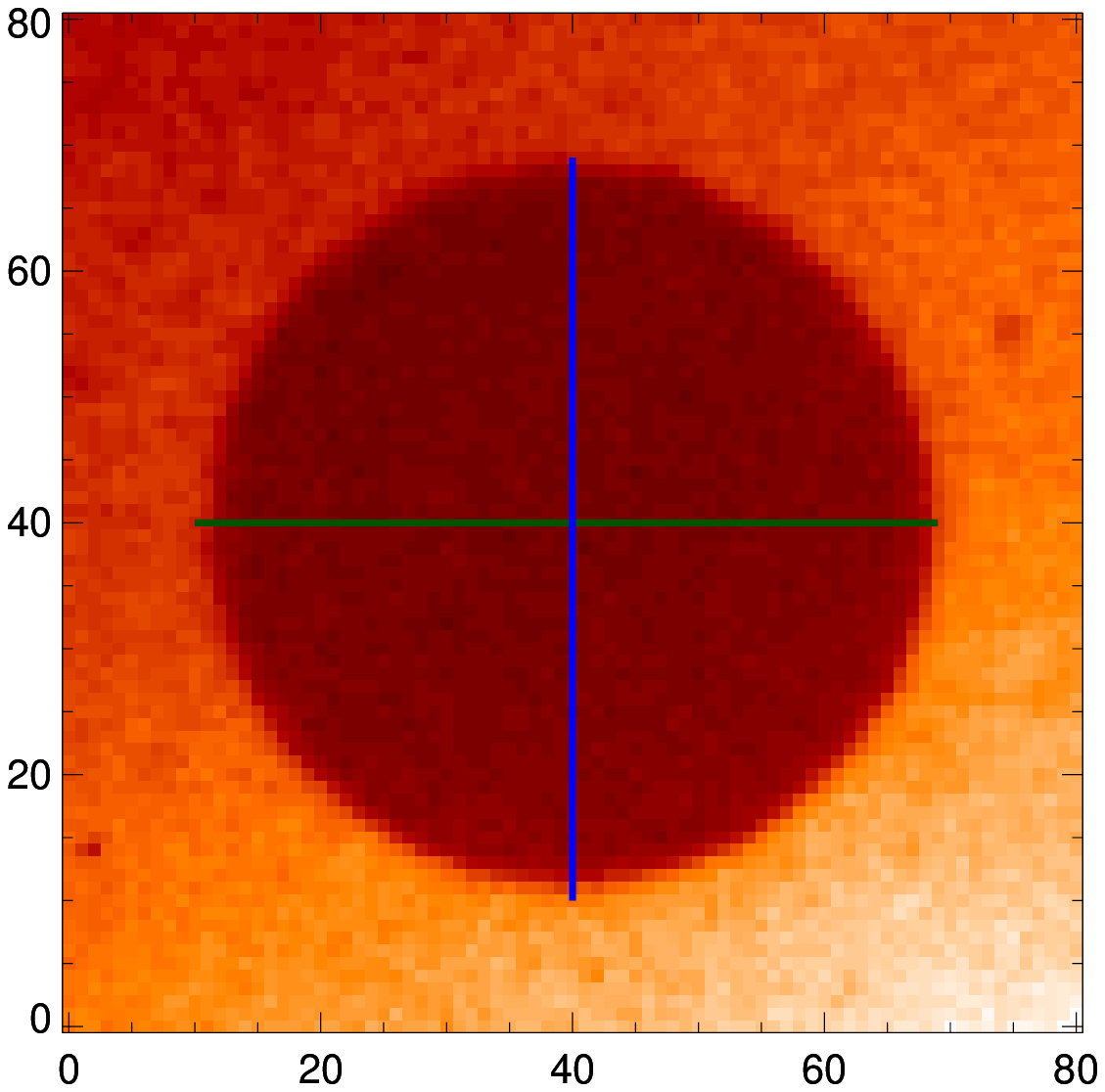}
 \hspace{10cm}
  \end{minipage}

 \begin{minipage}[b]{0.43\linewidth}
 \includegraphics[width=\textwidth]{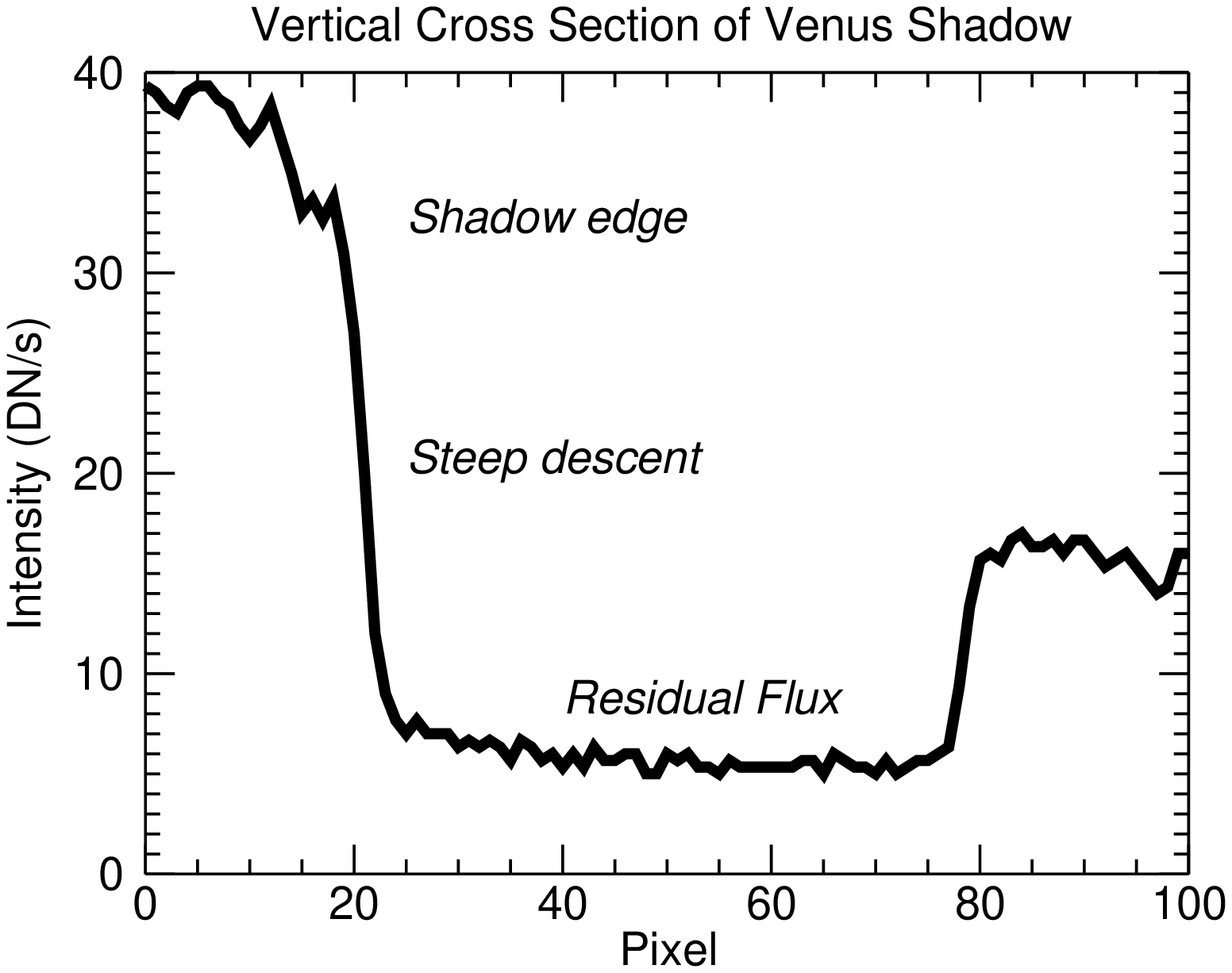}
  \hspace{0.1cm}
  \end{minipage}
  \begin{minipage}[b]{0.43\linewidth}
 \includegraphics[width=\textwidth]{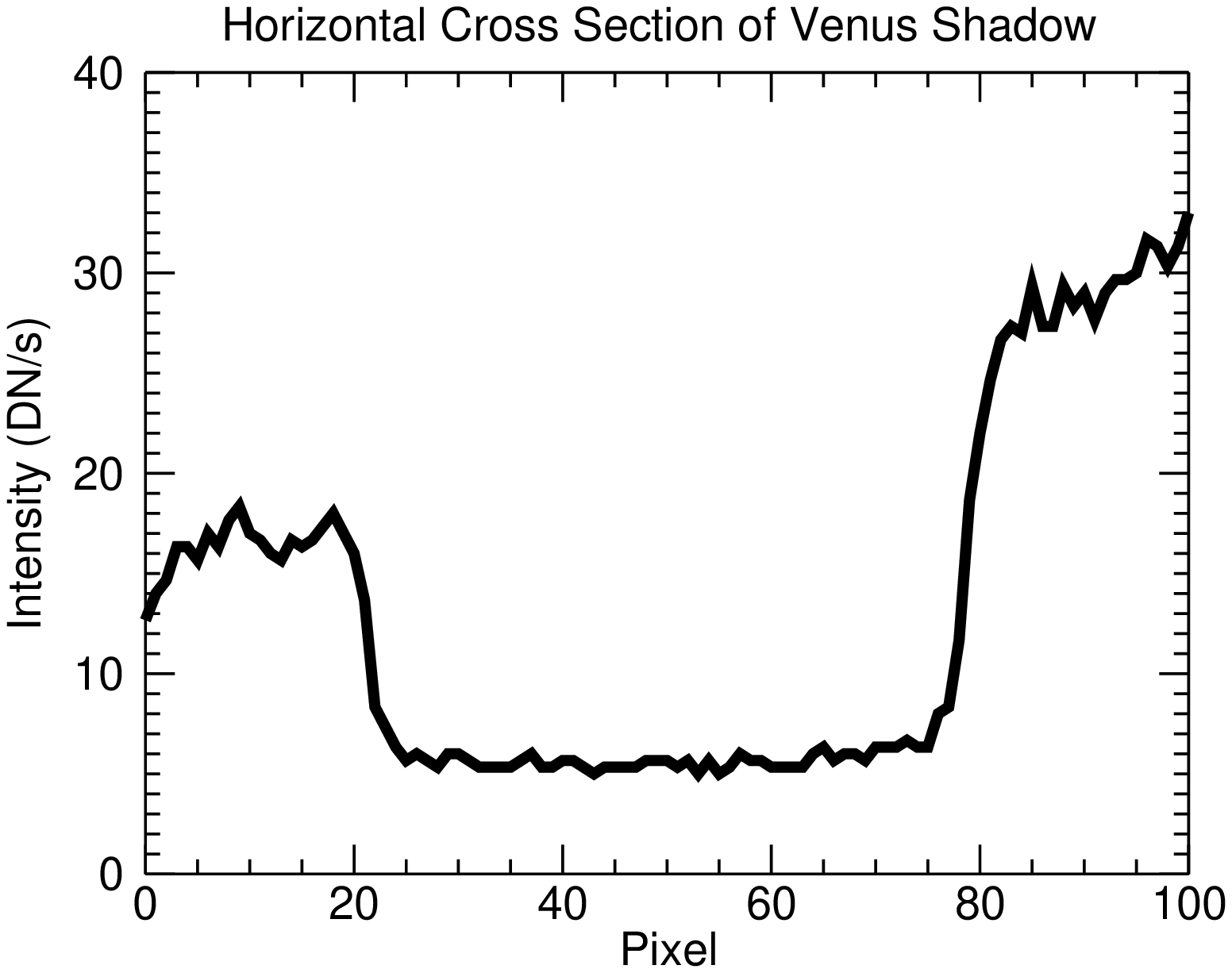}
   \hspace{0.1cm}
 \end{minipage}
 
 \caption{\small  Top Left: Venus transit above the active region.\newline
Top Right: Schematic view of horizontal (E-W) and vertical (N-S) strips, green and blue,
respectively.\newline
Bottom Left: Vertical IP of Venus\rq\  shadow. Bottom Right: Horizontal IP of
Venus\rq\  shadow. }
\label{VenusXcross}
\end{figure}

\begin{figure}[!h]           
\centering
\begin{minipage}[]{0.45\linewidth}
 \includegraphics[width=\textwidth]{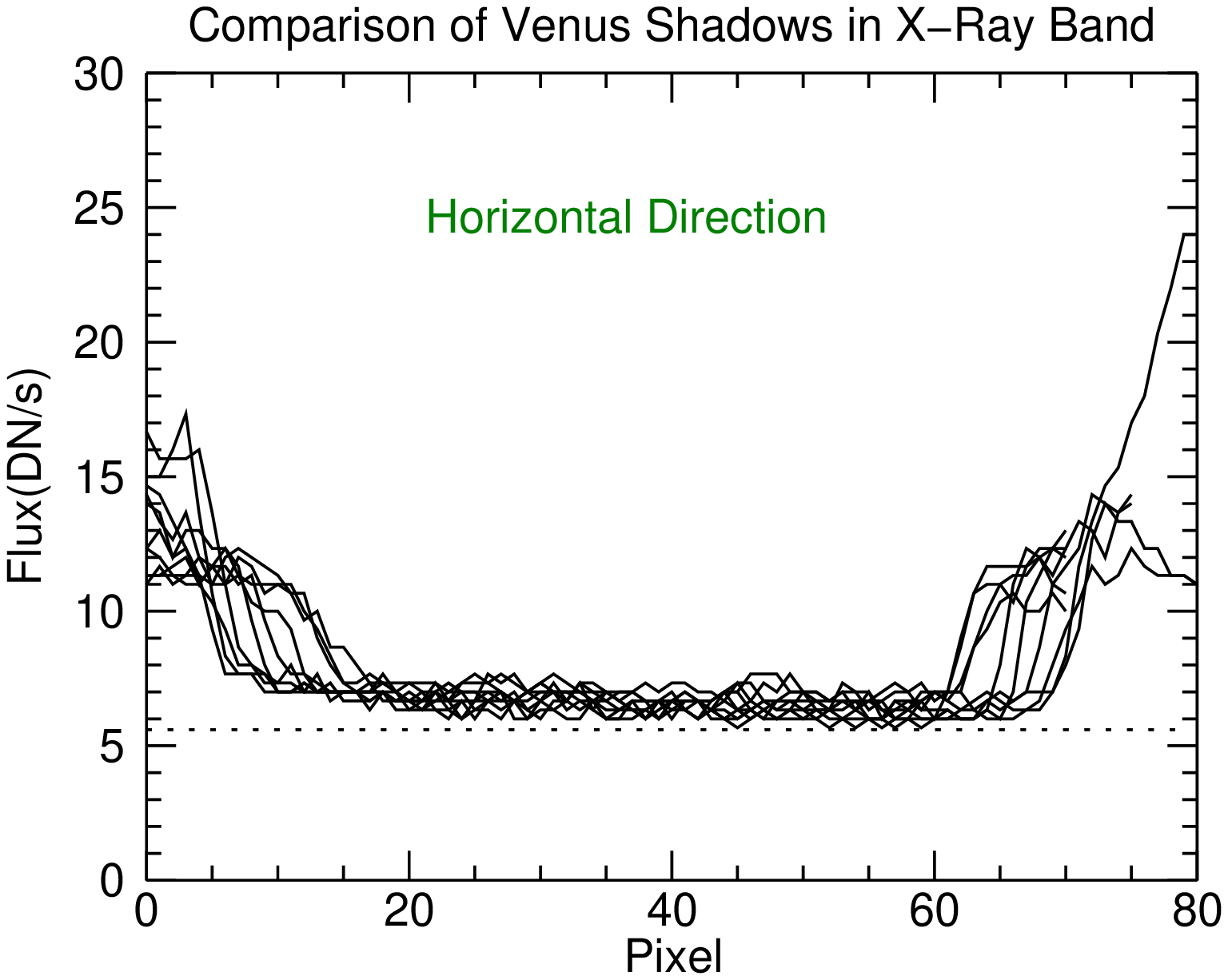}
  \end{minipage}
\begin{minipage}[]{0.45\linewidth}
 \includegraphics[width=\textwidth]{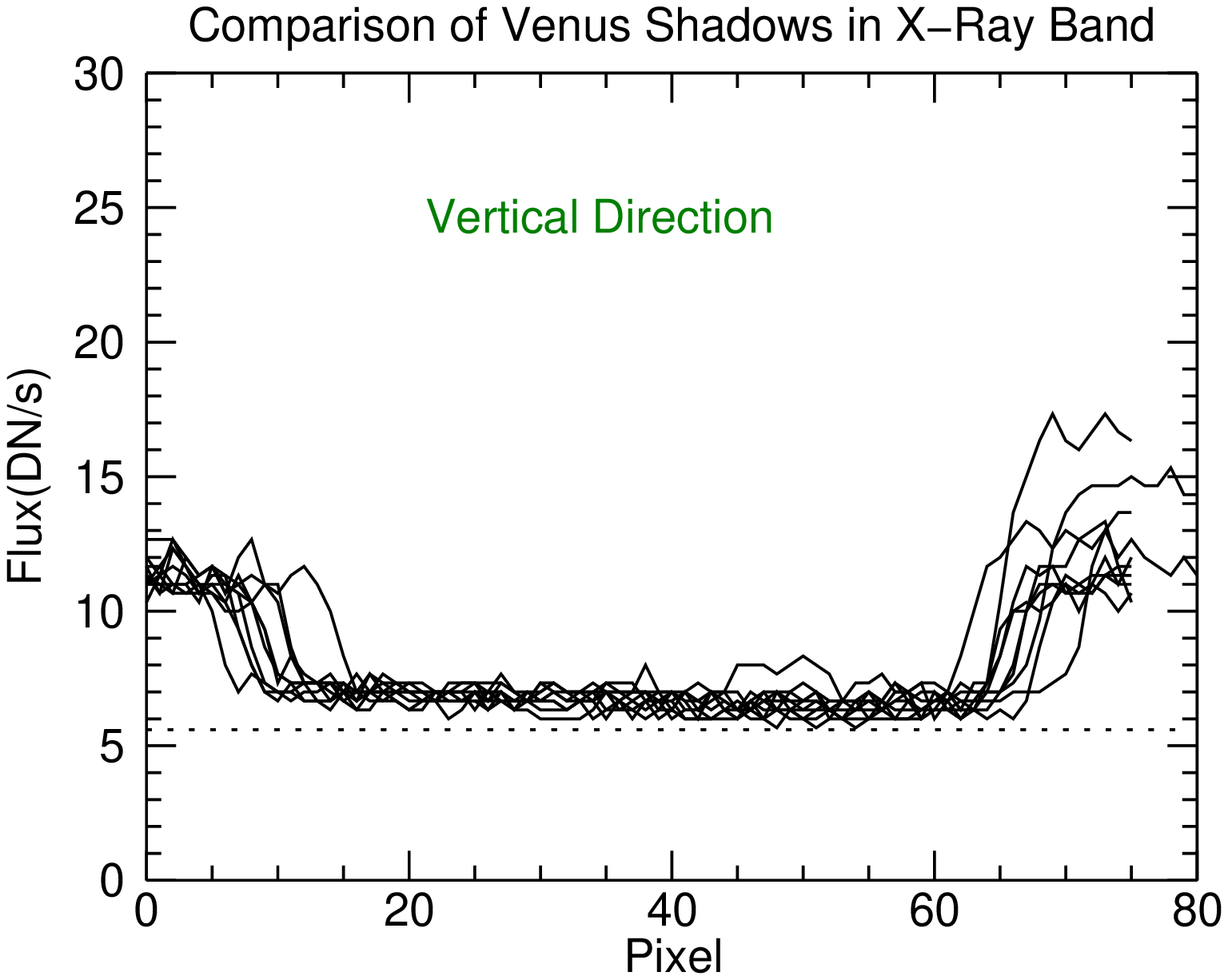}
  \end{minipage}
  
 \caption{\small IP of Venus\rq\  shadow in both horizontal (Left panel) and
vertical (Right panel) directions at different times of observations and
positions of Venus on solar disk.}
\label{VenusMnCrosses} 
\end{figure}

\begin{figure}[!t]               
\centering
  \subfigure[]{\includegraphics[trim={4.5cm 2cm 2cm 1cm},clip,width=4cm]{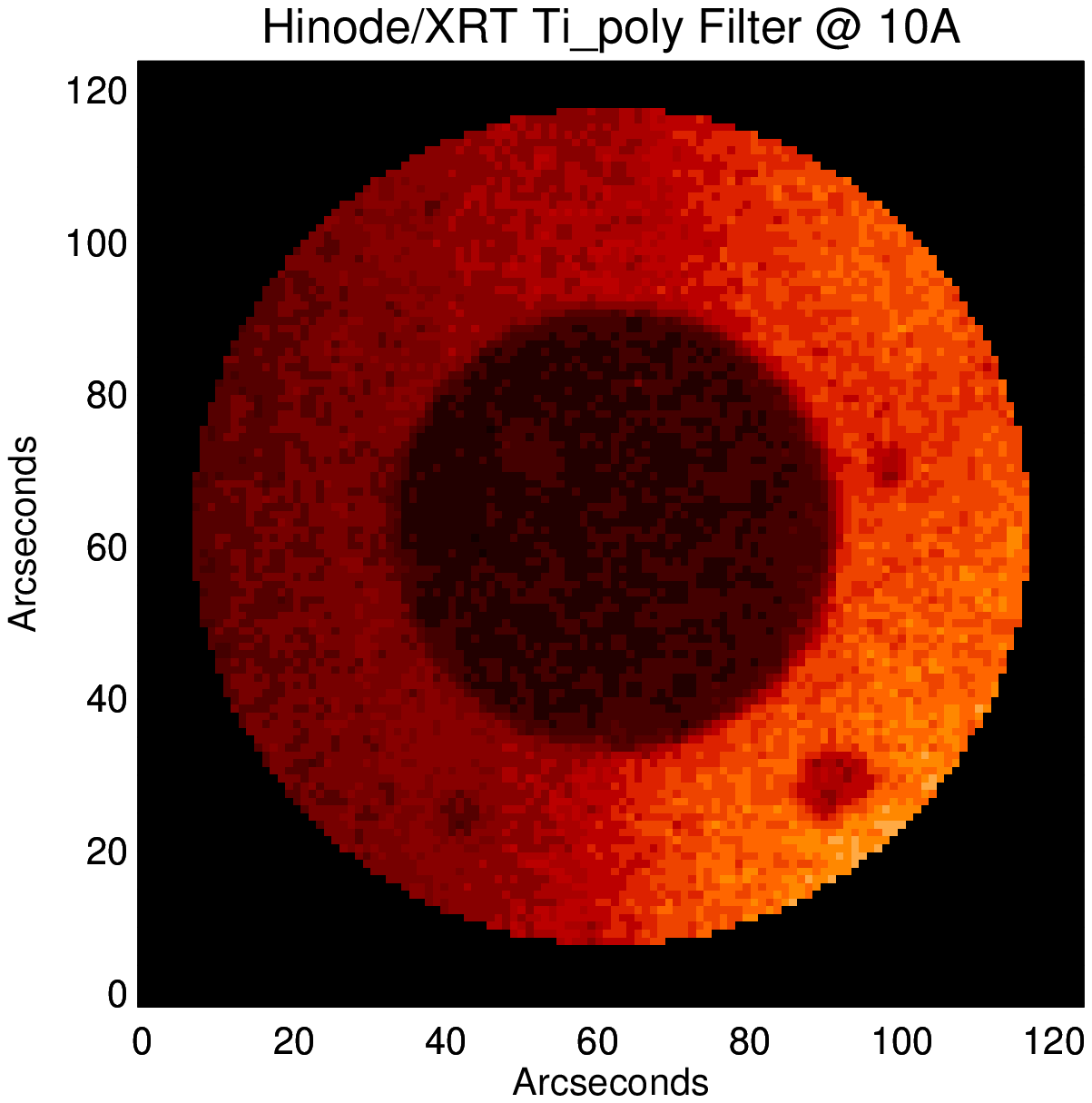}}
 \subfigure[]{\includegraphics[width=12cm]{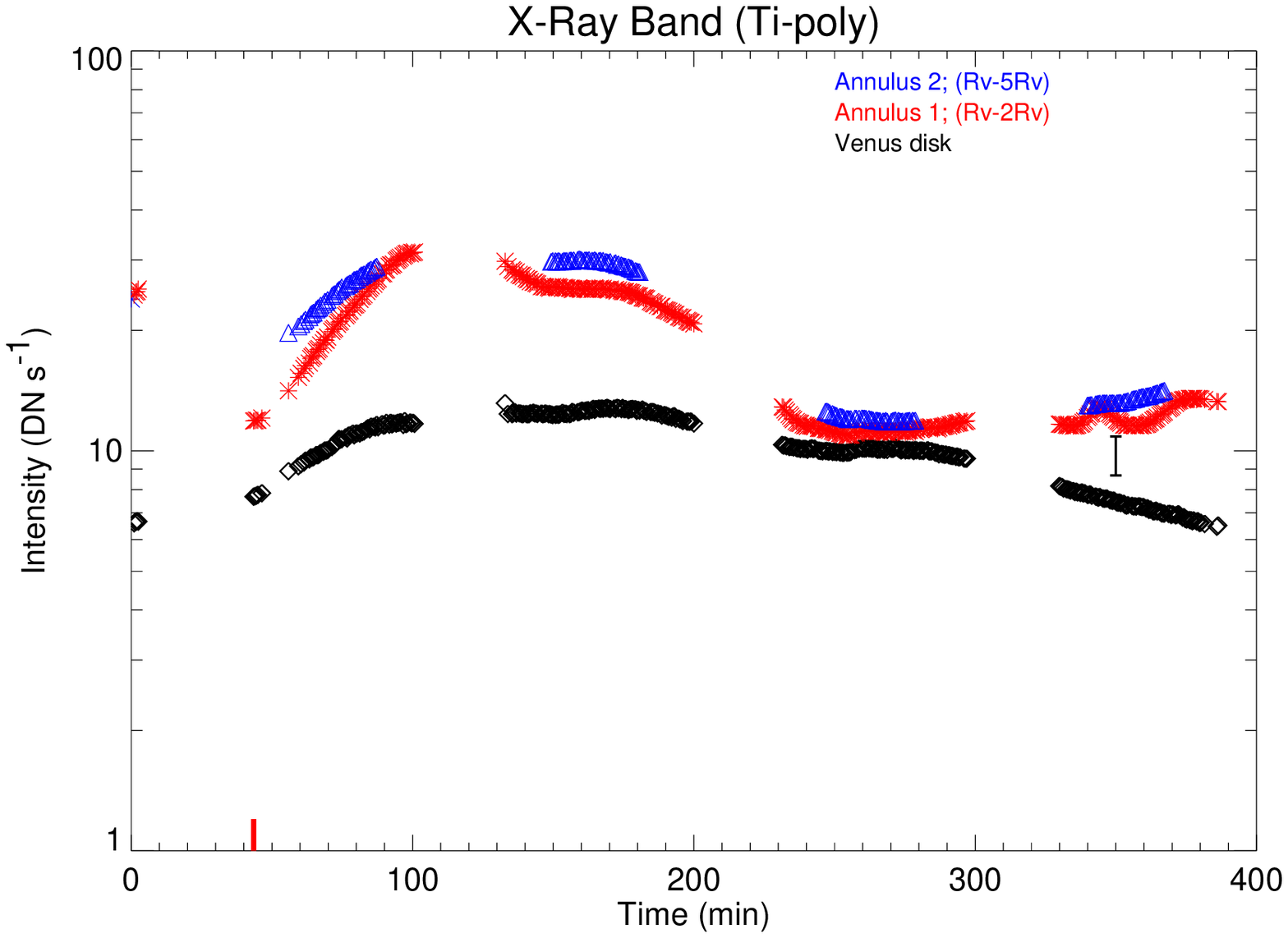}}

 \caption{\small Left: Central Black circle: Venus disk; Red  annulus: annulus 1
around the Venus disk.\newline
Right: the evolution of mean X-ray flux inside Venus disk (Black), annulus 1
(Red) and annulus 2 (Blue) vs. T$_{OBS}$.\newline
Annulus 1 has inner and outer radii R$_v$ and 2R$_v$. Annulus 2 has
inner and outer radii R$_v$ and 5R$_v$. The vertical bar on the right shows the typical
error size. The red vertical line in the lower left marks the first contact.}
 
\label{annulus}
\end{figure}

In Fig.~\ref{VenusXcross} we plot the Intensity Profile (IP) of Venus\rq\ 
shadow along both the horizontal and vertical directions in the X-Ray band, as
collected through the Ti-poly filter of XRT. Venus
casts a shadow with an angular diameter of $ \approx $ $60''$. The IP
of Venus\rq\  shadow consists of three parts: a shadow edge, a region of
steep descent on both sides and a residual flux.\\
The regions of steep descent have smooth corners on either side because
of the convolution of a step function with the PSF (\citealt{rea15};
\citealt{Web07}).\\ The X-Ray residual flux in Venus\rq\  shadow appears
too high to be compatible with background signal \citep{Kob14}. We have
superimposed in Fig.~\ref{VenusMnCrosses} the IPs taken at different times
and positions of Venus on the solar disk. We did not align the borders of
Venus, since the purpose here is only to show the level of residual flux
(albeit not sampling regularly the whole transit). \\
As we can see the level of residual flux is high at any time; the
intensity at the shadow\rq s edge strongly depends on the nearby (along
the line of sight) solar emission near Venus at the time the specific
frame was taken \citep{rea15}.\\
To check the effect of possible instrumental
scattering in XRT across Venus\rq\  shadow, especially when close to active
regions, we took the average flux measured in three regions: in the Venus
disk and in two concentric annuli around the Venus disk. Annulus 1 has
inner radius $R_v$, namely the Venusian radius, and outer radius 2$R_v$,
as shown in Fig.~\ref{annulus}. Annulus 2
has inner and outer radii $R_v$ and 5$R_v$, respectively. We plotted
the evolution of the mean flux inside each of these annular regions versus the time
of observation ($T_{OBS}$) in Fig.~\ref{annulus}, along with the flux
measured in Venus\rq\  shadow. In order to have a comparable time series in
all light curves we chose $T_{OBS} =$ 2012-06-05T21:58:39.912 of one {\it
Hinode}/XRT image as the reference time.  Also, the time of Venus\rq\  entrance
onto the solar disk is marked with a red vertical line. For each data point the
Poisson errors of DN (Digital Numbers) has been used as the error bars
in the light curves. This amounts to assume a DN-to-photon conversion factor of 1; according to
\cite{Nar11}  such a factor applies to $T \sim 1.5 $ MK, typical of the average, or quiet,
corona. The conversion factor changes only slightly over the temperature range of interest for the
non-flaring corona; since, also, the error depends on the square root of the photon number, the
error bar determined is adequate even considering the multi-temperature corona\\ 
The initial high annulus flux is due to limb brightening, crossed during
the initial phase of the Venus transit; then Venus gets close to a big active
region, during the central phase of transit, and the mean flux of both the
Venus disk and the annuli increases. (The maximum mean flux is measured in
this phase.) As Venus moves away from the active region the flux decreases
slowly. At the final stage, Venus completes the transit and
touches the other limb with a small increase in mean flux at the end of
all of the three curves. The blue curve does not cover the full data set:
for some images, the annulus with the outer radius 5$R_v$ extends beyond
the borders of the X-ray image.\\

\subsection{Mercury IP Analysis} 
\begin{figure}[!ht]         
\centering
 \subfigure[]{\includegraphics[width=8cm]{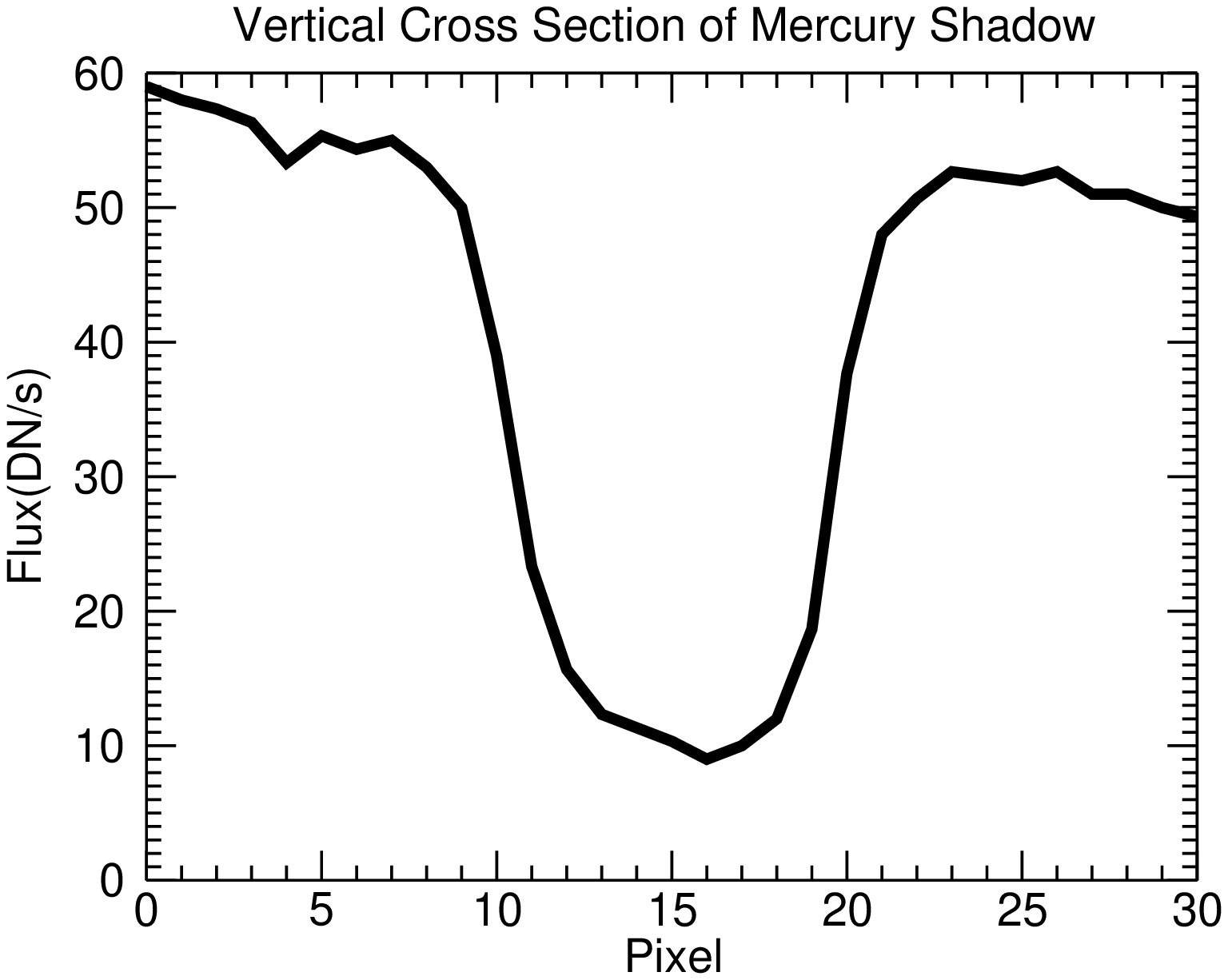}}
 \subfigure[]{\includegraphics[width=8cm]{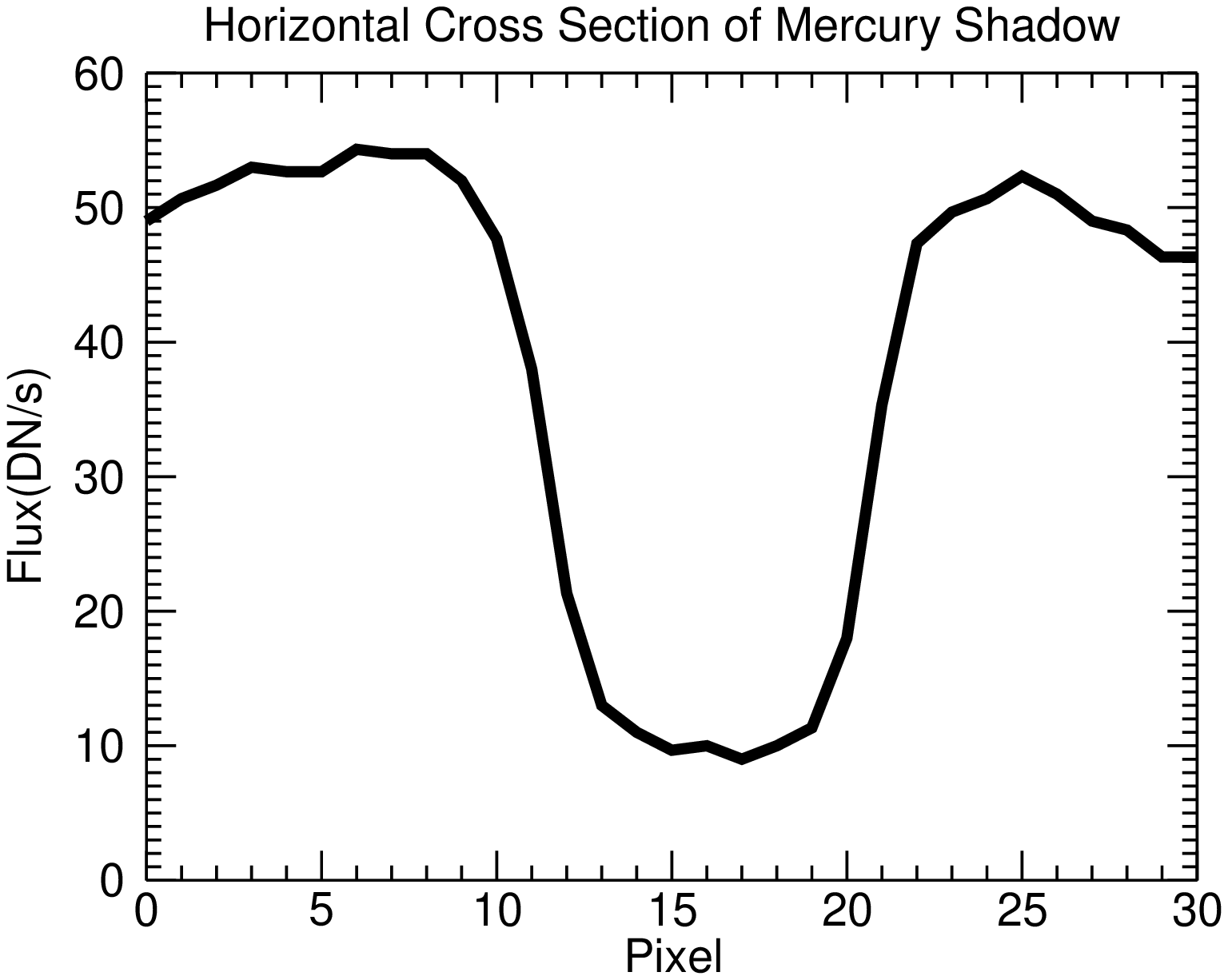}}
 
 \caption{\small Left: Vertical IP of Mercury\rq s shadow. Right: Horizontal
IP of Mercury\rq s shadow. Images taken through the Al-poly filter of XRT.}
\label{IPMercury}
\end{figure}

Since the
atmosphere of Venus may contribute to --- or be the cause of --- the residual
flux in IPs of Venus\rq\  shadow, we considered the shadows of other celestial
objects occulting the Sun but lacking an atmosphere, in order to remove the
possible effects of atmosphere.\\ As a first choice we selected Mercury,
already analyzed by \citet{Web07}. If some effect due to
PSF scattering is present in the case of Venus, it should be stronger in the case of the
smaller Mercury disk: Mercury casts a shadow with an angular diameter
of $\approx $ $10''$.\\
We have also made some analysis, not reported here, of the Moon's shadow
during solar eclipses observed with {\it Hinode}/XRT and found almost zero
signal coming from the Moon's X-ray shadow.
In Fig.~\ref{IPMercury} we have plotted the IP of Mercury\rq s shadow in the
X-Ray band, as taken through the Al-poly filter of XRT,
along both the horizontal and the vertical directions. The relevant
images were $384 \times 384$ pixels large.\\
\newline In the case of Mercury we initially find a residual flux, at a level
comparable to that in Venus\rq\  shadow, as well as a smooth profile. Therefore
the effect appears to be, at first sight, the same for Venus and
Mercury.\\
As a next step, in order to
remove possible instrumental effects due to the PSF, we deconvolved
Venus images using the \textit{Hinode}/XRT PSF and other codes,
and compared the relevant results. We also deconvolved Mercury images
with the same tools to cross-check the results.\\

\section{Deconvolution} 
Among different indirect methods of deconvolution
such as least-squares fit, Maximum Entropy, Maximum likelihood
\citep{Sta02}, and Richardson-Lucy (\citealt{Ric72}; \citealt{Luc74}), we used
the codes based on Maximum Likelihood (M-L) and Richardson-Lucy (AIA
Richardson-Lucy; AIA) available in SolarSoft 
IDL libraries. For a short description of the codes that we used, please
refer to Appendix B.\\
With the above codes and the \textit{Hinode}/XRT
PSF we performed deconvolution of the images, and then compared the results
to pinpoint similarities and differences; the Venus Ti-poly images were $
384 \times 384 $ pixels large. We repeated the cross section
analysis, presented before, for the deconvolved images.

\subsection{Deconvolution of Mercury shadows}  
\begin{figure}[!ht]              
\centering
 \subfigure[]{\includegraphics[width=8cm]{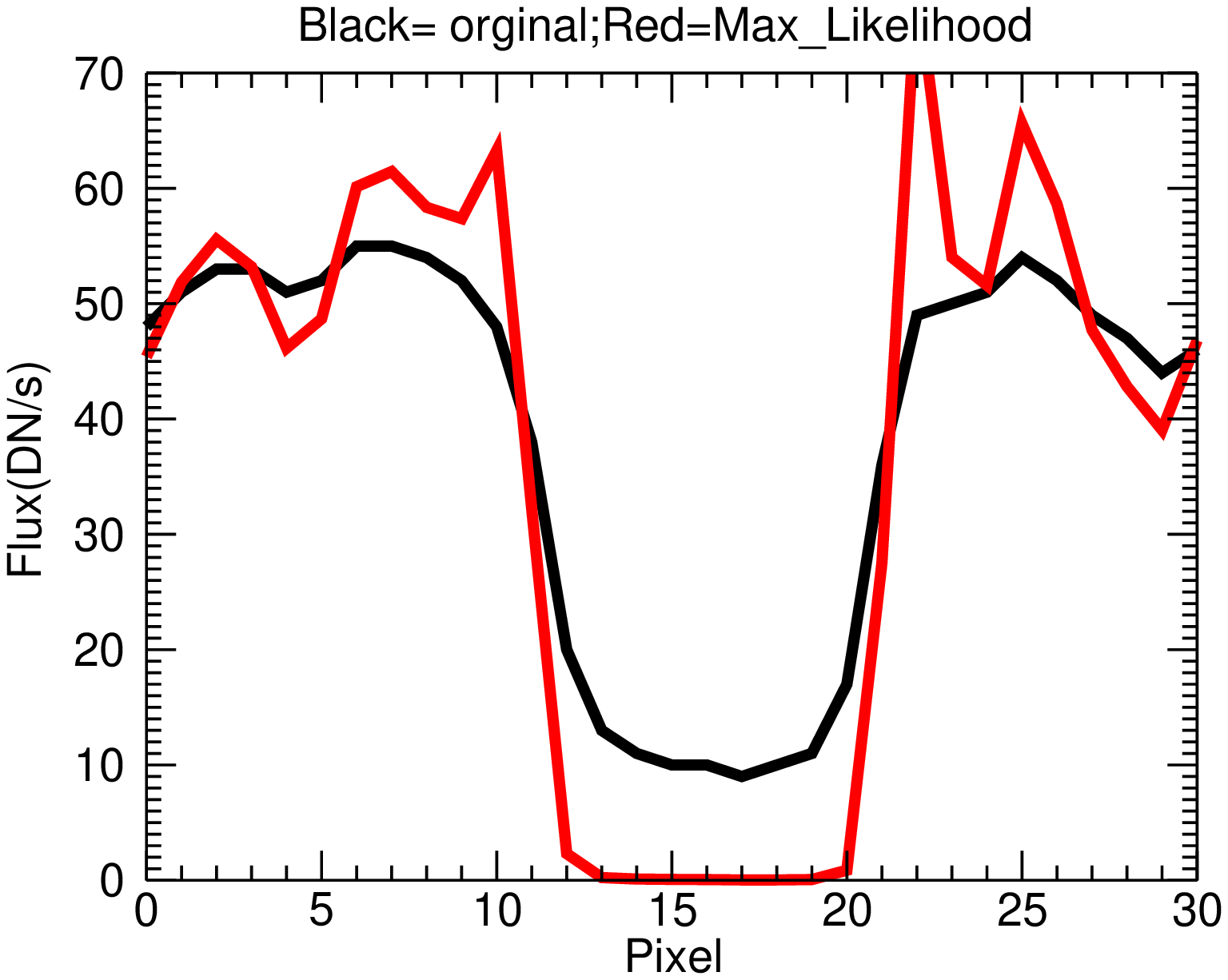}}
 \subfigure[]{\includegraphics[width=8cm]{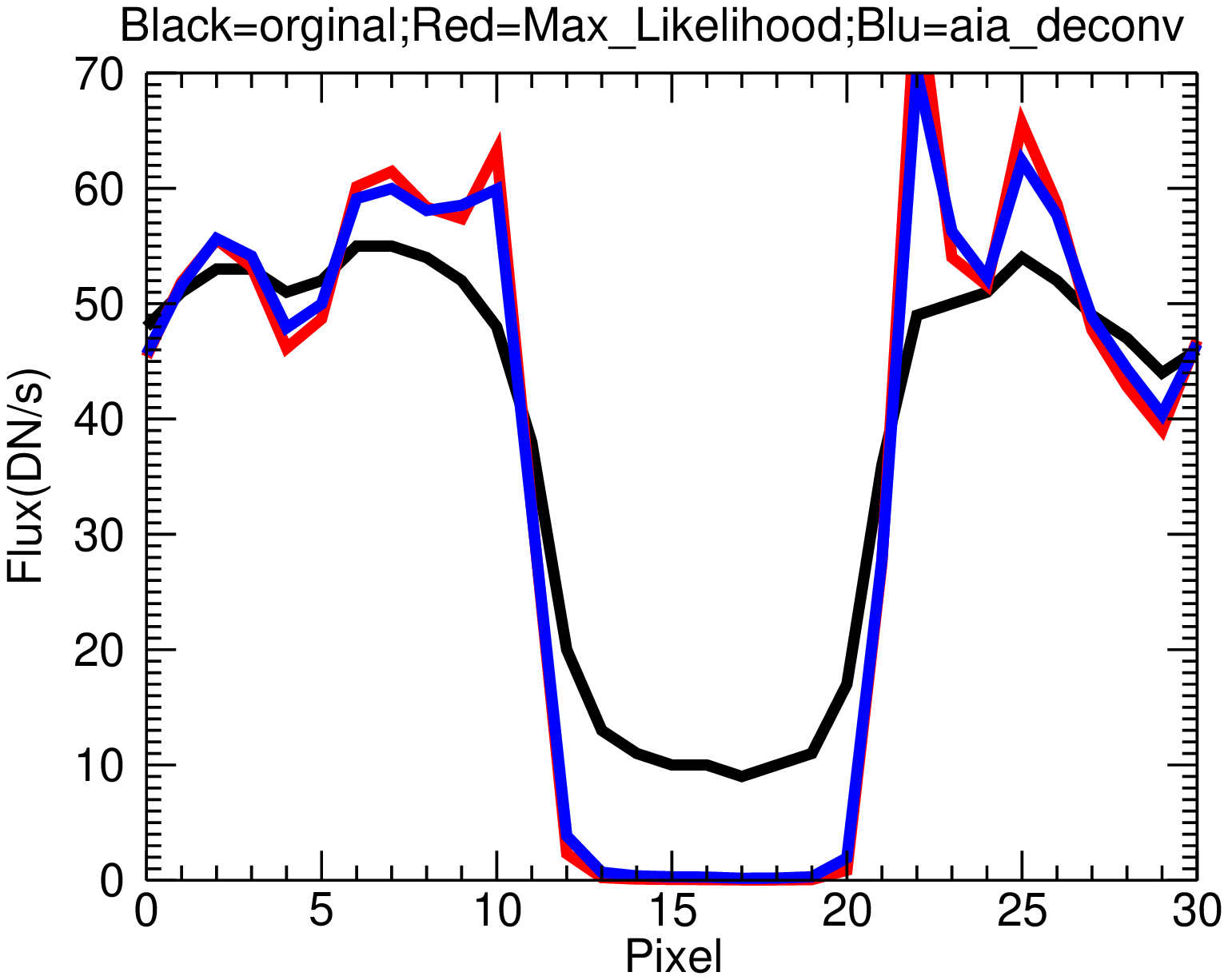}}
 
 \caption{\small Left: Mercury IP before (Black) and after deconvolution
with the M-L code (Red). \newline
Right: comparison between IP before (Black) and after deconvolution 
with AIA (Blue) and M-L (Red). }
\label{MercuryDecon}
\end{figure}

Mercury and Venus have been observed at different times, in 2006 and in
2012 respectively, and with different filters. However our aim here is
just to check the performance of the updated PSF in removing any emission
concerning the instrumental scattering.  The cross sections of Mercury
shadow, before and after deconvolution, are presented in 
Fig.~\ref{MercuryDecon}. After
deconvolution, the cross section of Mercury\rq s shadow has practically zero
residual flux with edges sharper than those of the original profiles.\\
 \newline These results are very important since they not only confirm the accuracy
of the updated  \textit{Hinode}/XRT PSF but show that, at least in the
case of Mercury, the residual flux is due to PSF scattering. An analogous
study was done by \citep{Web07}, with similar results, using a previous
version of the PSF of \textit{Hinode}/XRT.\\

\subsection{Deconvolution of Venus shadows}  
\begin{figure}[!htb]     
\centering
 \subfigure[]{\includegraphics[width=8cm]{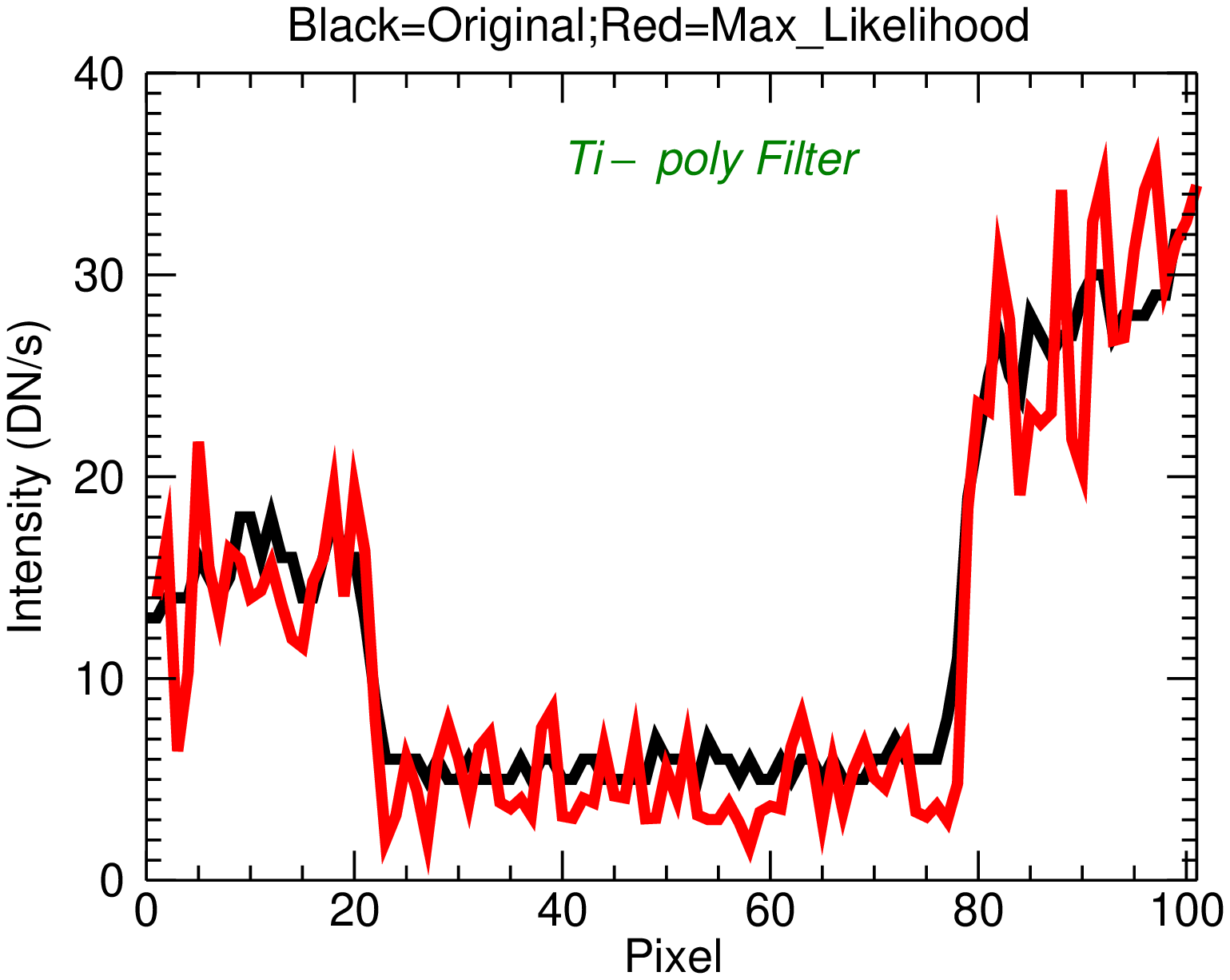}}
 \subfigure[]{\includegraphics[width=8cm]{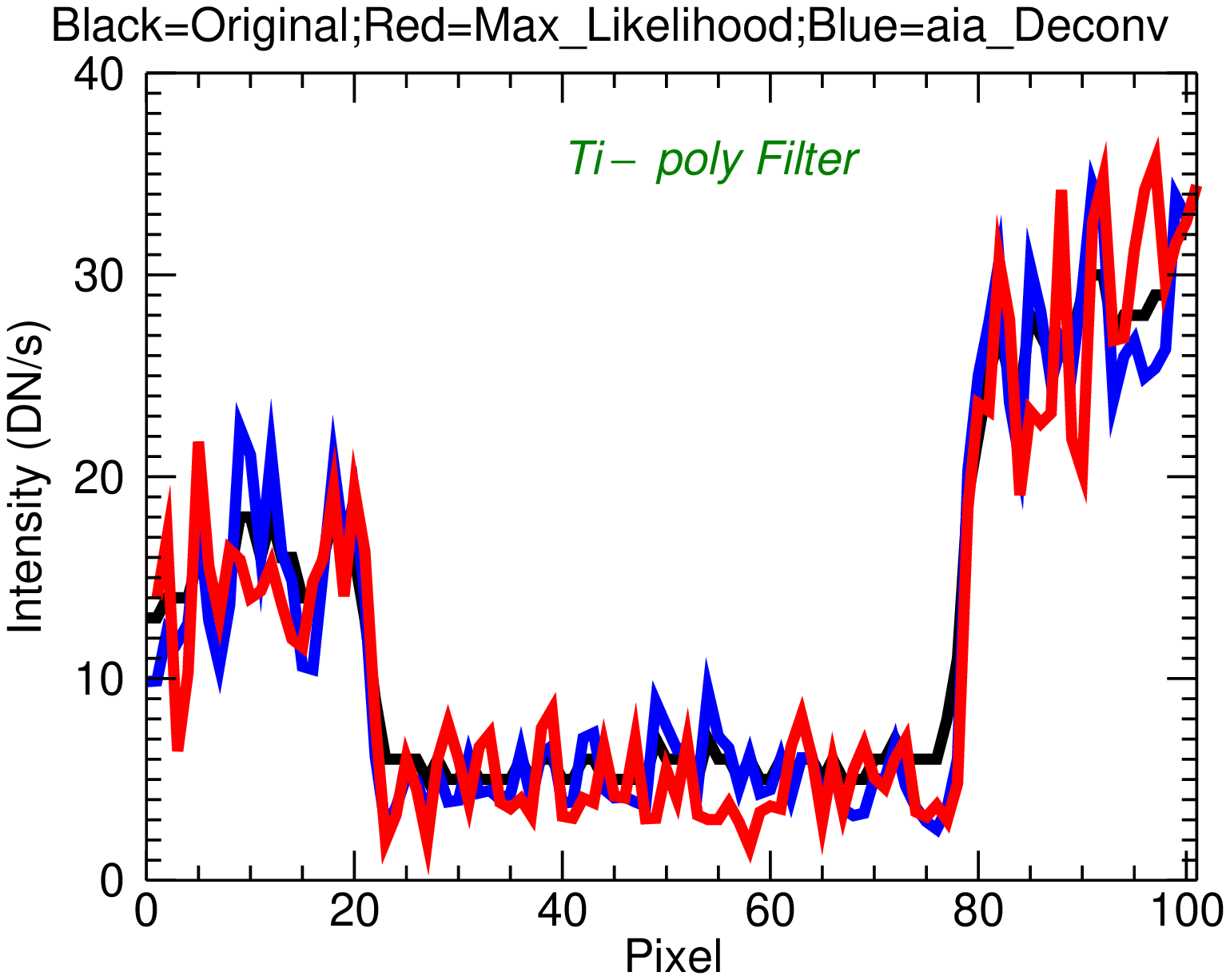}}
 
 \caption{\small Left: Venus IP before (Black) and after deconvolution made with the M-L code (Red). \newline
Right: Comparison between IP before (Black) and after deconvolution made
with AIA (Blue) and M-L (Red).
 }.
\label{VenusIPdeconv}
\end{figure}

Cross sections of Venus\rq\  shadow after deconvolution are shown in
Fig.~\ref{VenusIPdeconv}.
 \newline These cross sections of Venus images deconvolved with
the M-L and AIA codes show that:
\begin{description}
 
\item[$\bullet$]In some cases the cross sections of images deconvolved 
with the M-L code have more fluctuations in comparison to those obtained with
the AIA code;
\item[$\bullet$]For Venus, similarly to the Mercury case, the borders seem to be sharper
after deconvolution;
\item[$\bullet$]Residual flux is present in Venus images even after
deconvolution; such a flux is significantly higher than the noise.
\end{description}

Residual flux present in Venus cross sections after deconvolution does not
appear to be due to the \textit{Hinode}/XRT PSF, since the accuracy of the PSF
has been confirmed in the Mercury analysis. Being that the angular size of
Mercury is
considerably smaller than that of Venus, any effect of PSF scattering
should manifest itself more in the Mercury cross sections. \\
Since both the M-L and AIA codes are iterative we changed the number of
iterations during the deconvolution process for some images in each dataset to
check the effect of iteration, especially to see whether increasing
the number of iterations led to the residual flux being further decreased or
removed. The
trend is that with increasing the number of iterations the residual
flux is still present and its mean value for any reasonable number of iterations
is very constant,
except that with increasing iterations the
fluctuations increase in the IPs. Generally the M-L code is more sensitive to
noise and the quality of the images. \\
So the presence of a significant residual flux in Venus\rq\  shadow is not due
to instrumental scattering but should be related to Venus; for instance,
it could originate from some effect occurring in Venus\rq\  atmosphere. \\
Comprehensive analysis of deconvolutions show that:

\begin{description}

\item[$\bullet$] The AIA code does not conserve the total flux, yielding
curves with 15\% of total flux, so for each image we readjusted the
amplitude to conserve the total flux;

\item[$\bullet$]Deconvolution causes artifacts and spurious ``spikes"
at the edge (borders), a common problem in deconvolution which, in the
case of Venus, are well identified and do not affect the evaluation
of the average flux in the shadow (cf.\ Fig.~\ref{VenusIPdeconv}).

\end{description}

\begin{figure}[!htb]       
\centering
\includegraphics[scale=0.8]{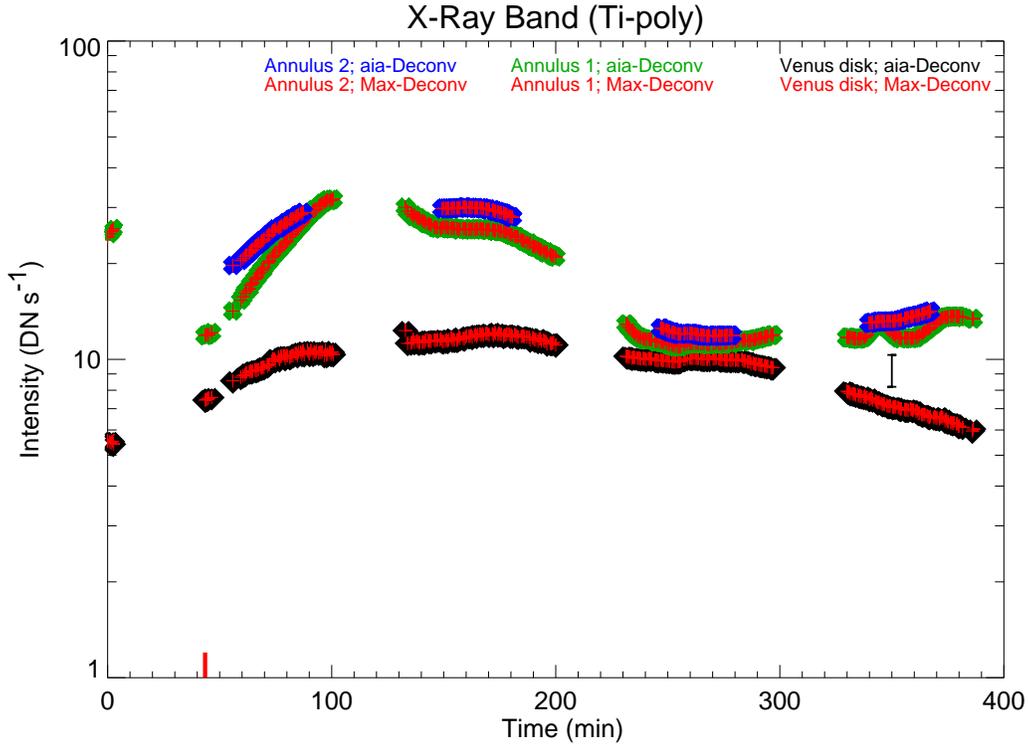}
 
 \caption{\small Evolution of mean Ti-poly flux after deconvolution: inside Venus disk
with AIA (Black) and M-L (Red) codes; inside 
annulus 1 
with AIA (Green) and M-L
(Red) codes; inside annulus 2
with AIA (Blue) and M-L (Red) codes. The bar on right shows typical
error sizes. The red vertical line in the lower left marks the first contact.}

\label{LCTiDecon}
\end{figure}

We again followed the evolution of the flux in Venus\rq\  shadow and in two
reference annuli, as done in Section 2, after deconvolution. We plotted
the evolution of mean flux inside each of these three regions versus
T$_{OBS}$ in Fig.~\ref{LCTiDecon}.
The space-averaged fluxes obtained after deconvolution with the two
methods are virtually the same resulting in three light curves, each
being two superimposed.\\
The most important points in Fig.~\ref{LCTiDecon} are: 
\begin{description}
\item[$\bullet$]The amount of mean flux inside Venus\rq\ disk after
deconvolution has decreased slightly, especially where close to the active region,
therefore deconvolution appears to have removed just a small amount of X-ray
flux from Venus\rq\  shadow;
\item[$\bullet$]The mean fluxes inside the two annuli have not changed after
deconvolution (cf.\ also Fig.~\ref{VenusIPdeconv});
\item[$\bullet$]We see that the flux inside Venus\rq\  shadow and that inside
the two annuli gradually rise as Venus gets more and more inside the solar
disk and decreases thereafter; however the flux inside
Venus\rq\  shadow is not strictly correlated to that inside the two annuli.
\item[$\bullet$]There may be some relationship between the observed residual
flux and the high surrounding flux;
\item[$\bullet$]The mean flux value for both the AIA and M-L deconvolution codes
are virtually the same despite the fact that the profiles for the deconvolved images are
not the same.\\

As an additional test on the error, we have determined the standard deviation of
the residual flux values inside Venus disk 
and found that it varies
between 0.4--1.0 DN s$ ^{-1} $, which is negligible in comparison to the observed residual
flux ($>$ 5 DN s$ ^{-1} $).\\

\end{description}

\section{Light Leak Contamination}  
\subsection{ Light Leak Effect on XRT Filters} 
An increase in XRT's straylight was detected on May 9th of 2012, shortly before the 
Venus transit (5th--6th June 2012), which causes significant visible 
light contributions to the X-Ray images in some filters. In addition, a sudden
increase of intensity by a factor of 2 was observed in the visible light
measurements 
(i.e., in the G-band channel). At the same time, the XRT team recognized wood-grain like 
stripes in daily images taken with the Ti-poly filter \citep{Tak16}. The team believes the increase of visible straylight to have been caused by a pinhole puncture in the entrance aperture filters. \\
The analysis showed that the light leak affects only some of the X-Ray
filters: a minor effect was detected for the Al-mesh and Al-poly filters but
it was very small ($\le ~5$ DN s$ ^{-1} $), while it strongly affected the Ti-poly and C-poly filters \citep{Tak16}.\\
In order to exclude the possibility that Venus residual
flux in Ti-poly could be due to the straylight, we used data collected
with the Al-mesh filter to repeat the analysis. Importantly, the light leak
has a very small effect on the Al-mesh filter, to such a level that it can be neglected
(\citealt{Tak16}).

\subsection{Al-mesh Filter Analysis} 
\begin{figure}[!h]  
\centering
 \subfigure[]{\includegraphics[width=8cm]{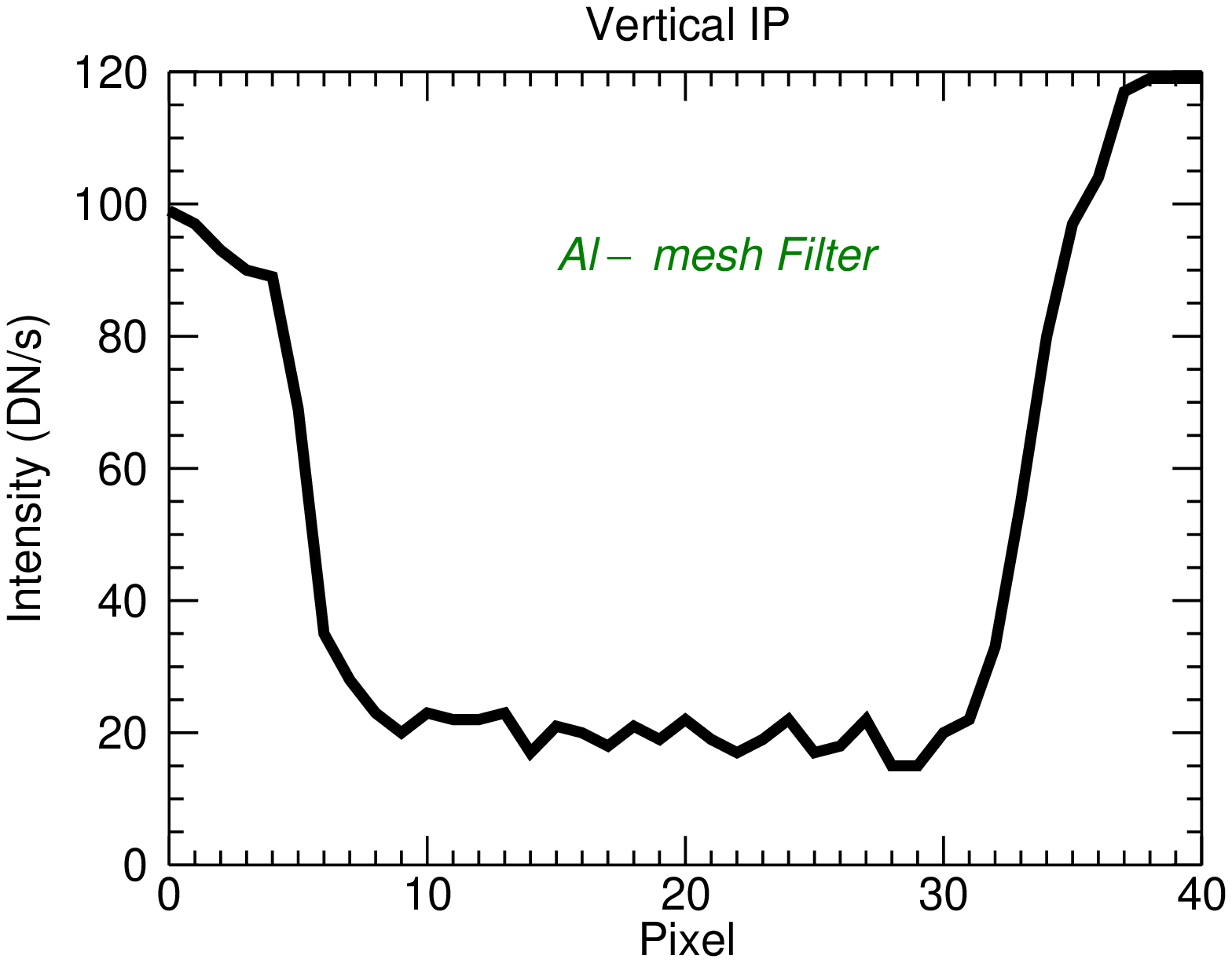}}
 \subfigure[]{\includegraphics[width=8cm]{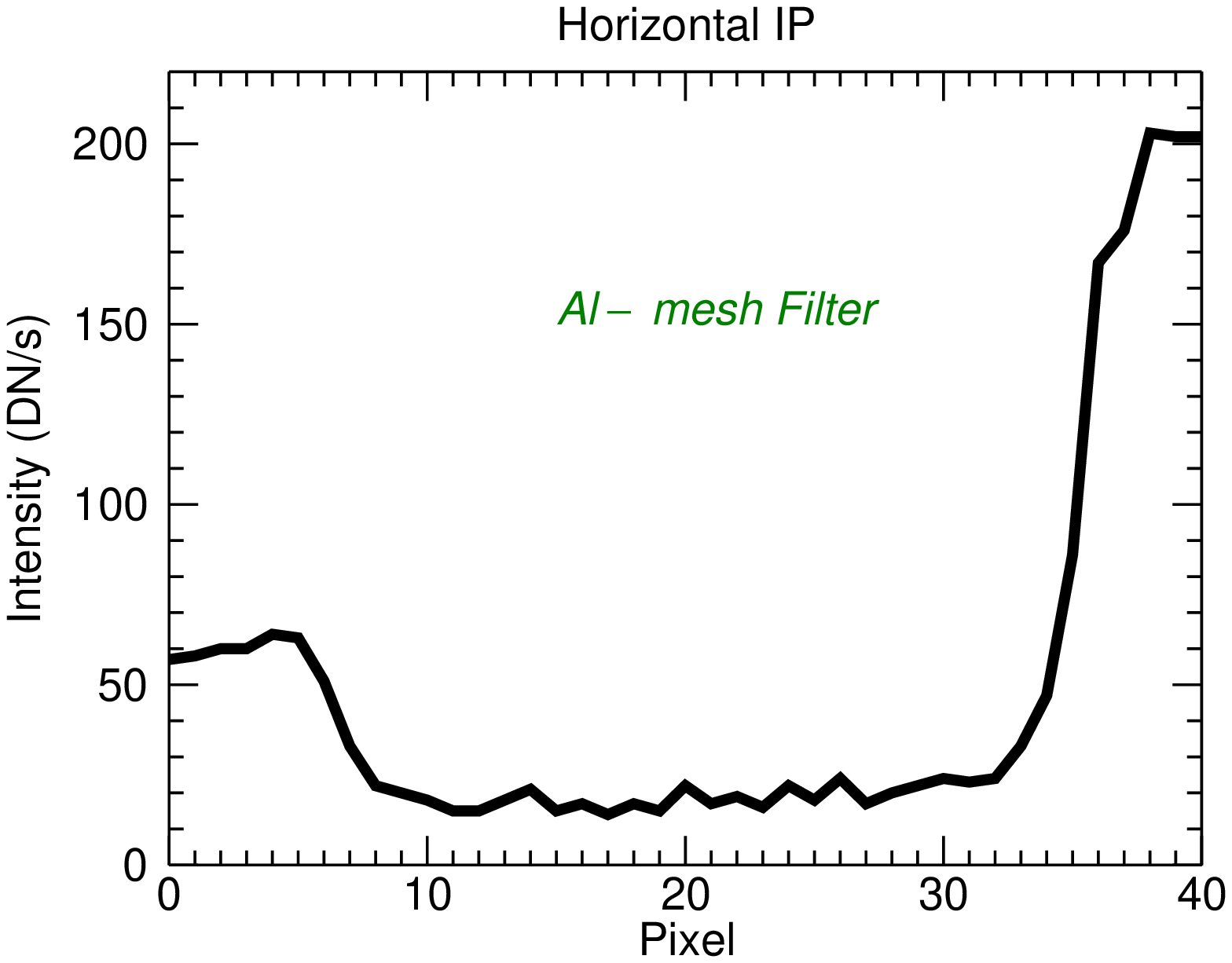}}
 \caption{\small Left: Vertical IP of Venus\rq\  shadow. Right: Horizontal IP of
Venus\rq\  shadow. (XRT Al-mesh filter)}.
\label{IPAL}
\end{figure}

In Fig.~\ref{IPAL} we present a typical IP of Venus\rq\  shadow in both
horizontal and vertical directions, taken in an image collected with the Al-mesh
filter.\\
 \newline 
As we can see:
\begin{description}

\item[$\bullet$]The residual flux is still present in all IP plots.
\item[$\bullet$]The intensity profiles of the Al-mesh filter appear
approximately 3-5 times higher than Ti-poly ones; the reason is that
Al-mesh images are binned $2 \times 2$  while Ti-poly data are binned $1
\times 1$ and the
filters have different transmissivity; the Al-mesh images are $ 192 \times
192$ pixels large.

\end{description}

\begin{figure}[!htb]             
\centering

 \subfigure[]{\includegraphics[width=8cm]{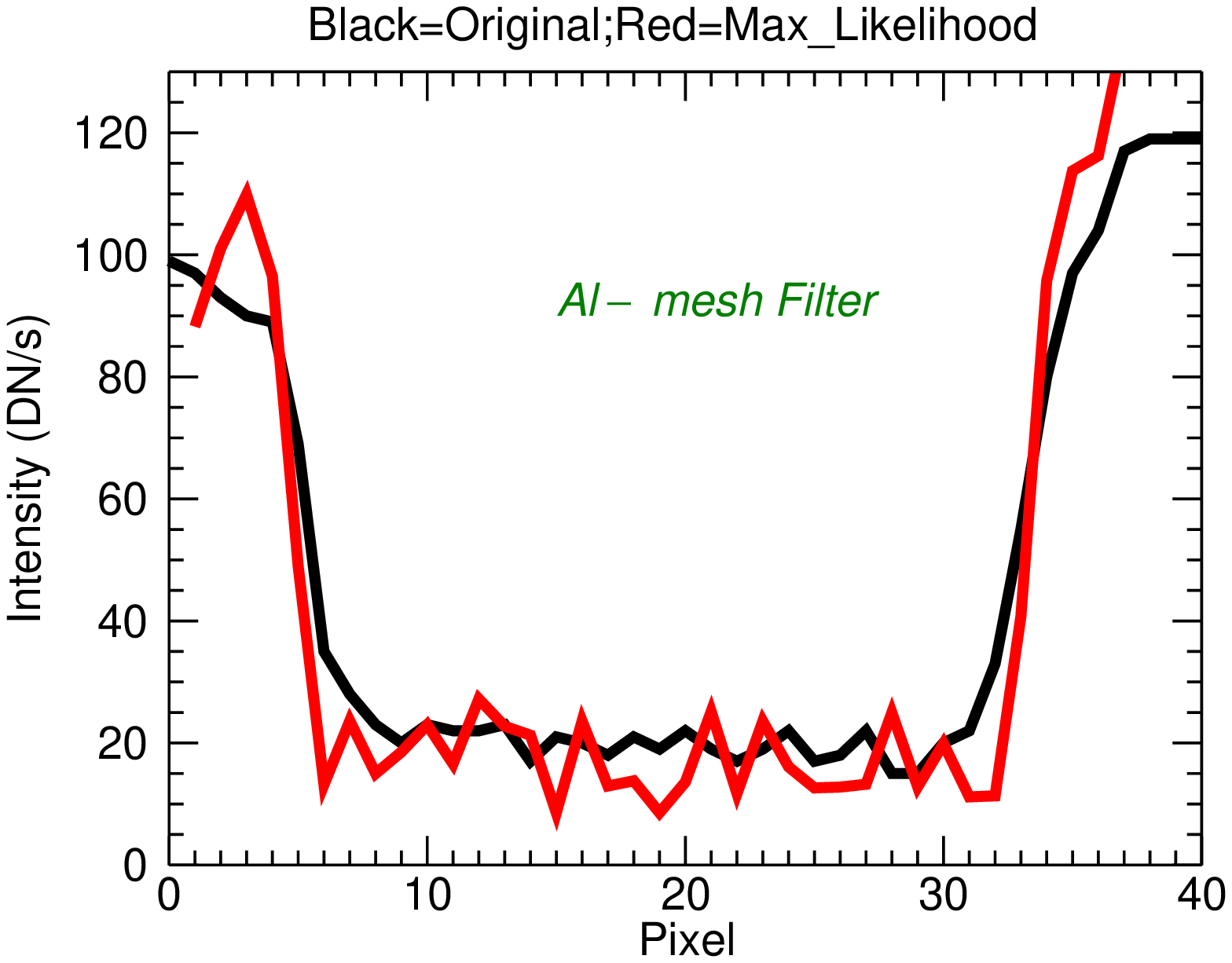}}
 \subfigure[]{\includegraphics[width=8cm]{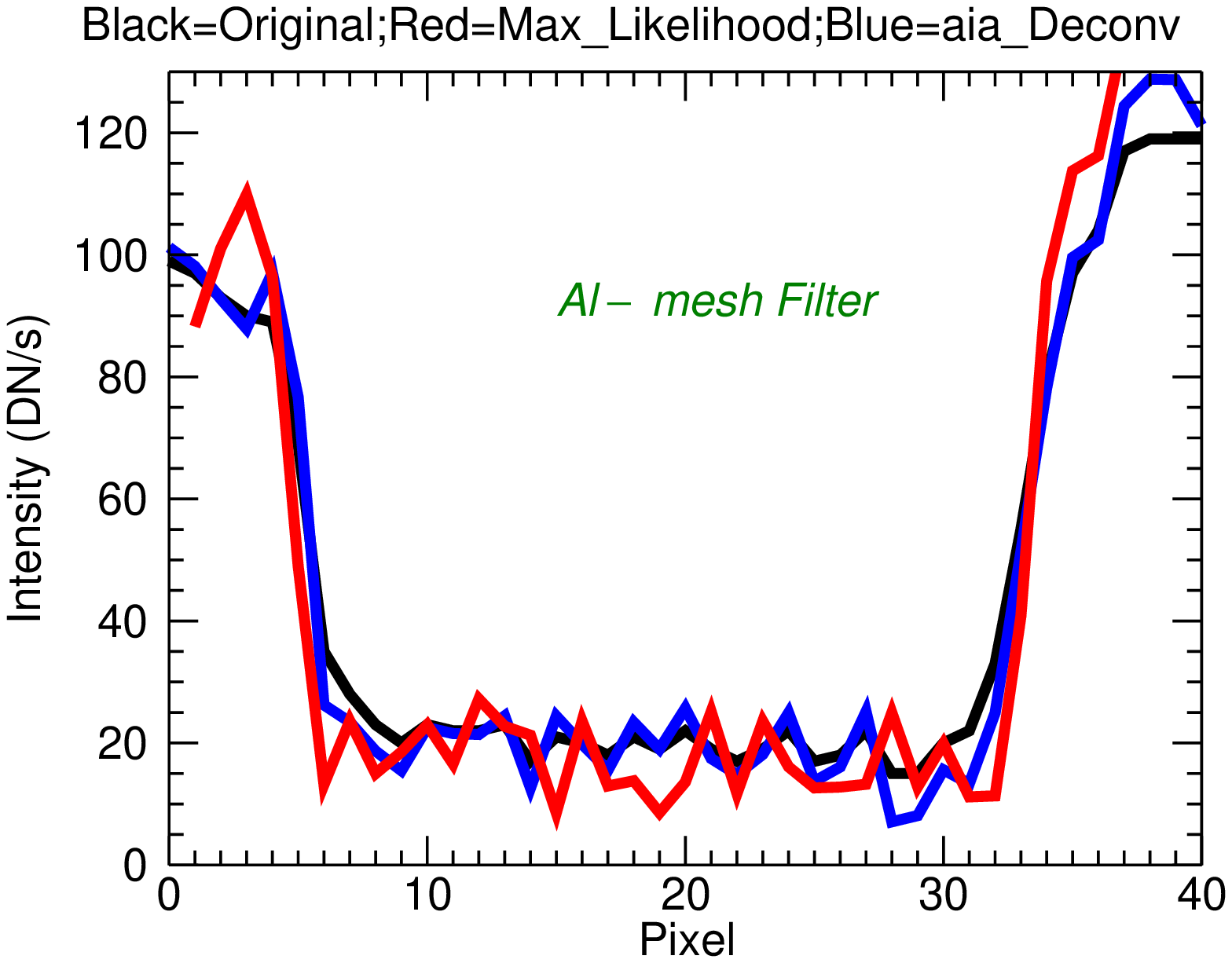}}

\caption{\small Left: Venus IP before (Black) and after deconvolution
with M-L code in the vertical direction (Al-mesh filter).\newline
Right: comparison of IP before (Black) and after deconvolution with AIA
(Blue) and M-L (Red) in vertical direction.}
\label{DeconIPAL}
\end{figure}

Also for Al-mesh data we deconvolved images to remove any effect due to
the PSF scattering. Sample IP results, after deconvolution, are shown for the
vertical direction in Fig.~\ref{DeconIPAL}.\\
Deconvolution analysis shows that:
\begin{description}
\item[$\bullet$]Artifacts and spurious ``spikes" at the edges (borders)
of the IPs are much stronger in Al-mesh images in comparison to Ti-poly images.
\item[$\bullet$]The AIA code does not conserve the total flux, yielding curves
with 60\% of total flux (in Ti-poly, 15\%), so for each image
we rescaled the amplitude to conserve the total flux.
\item[$\bullet$]The most important fact is that even after deconvolution
residual flux is still present in all of the IPs and is significantly higher
than the noise.
\end{description}

\begin{figure}[!t]    
\centering
\includegraphics[scale=0.8]{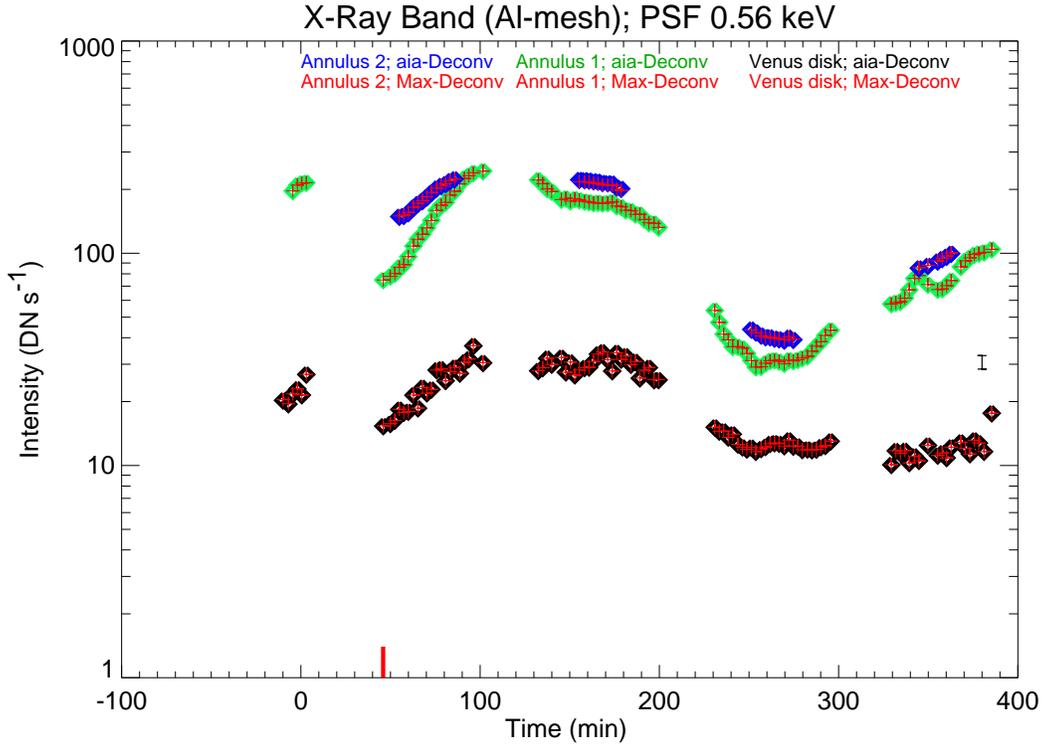}
 
\caption{\small Evolution of mean X-ray flux (as measured through Al-mesh
filter) inside Venus disk after deconvolution with AIA (Black) and M-L
(Red) codes, inside annulus 1 after deconvolution with AIA (Green) and
M-L (Red) codes, and inside annulus 2 after deconvolution with AIA
(Blue) and M-L (Red) codes. A typical error bar is shown on the
right. The red vertical line in the lower left marks the first contact.}

\label{LCDeconAl}
\end{figure}

We have also determined the evolution of the flux in Venus\rq\  shadow after
deconvolution for Al-mesh images and in two reference annuli,
similarly to what we have done for the Ti-poly data. Fig.~\ref{LCDeconAl} shows
the evolution of the mean flux inside each of these regions versus
T$_{OBS}$. \\
Also for Al-mesh data the DN to photon conversion factor of 1, used to derive the error bars,
is appropriate to $T \sim 1 $ MK and changes slowly over the T range of interest for the
non-flaring corona.\\
The ratio of the maximum value of flux inside annulus 1 to the lowest
one for Al-mesh data
is slightly more than 5, on the average. The ratio is different from that of
Ti-poly ($\approx 3$) probably because the light leak effects (if any) are very small in the case of Al-mesh.\\
We can safely state that the flux detected in Venus' dark side
most likely is not due to PSF scattering, noise or light leak, but it
may originate from some phenomenon related to Venus.\\

\section{The EUV and UV Flux Analysis} 
\begin{figure} [!htb]               


\centering
\includegraphics[trim={4.7cm 2cm 3cm 1.2cm},clip,scale=.75]{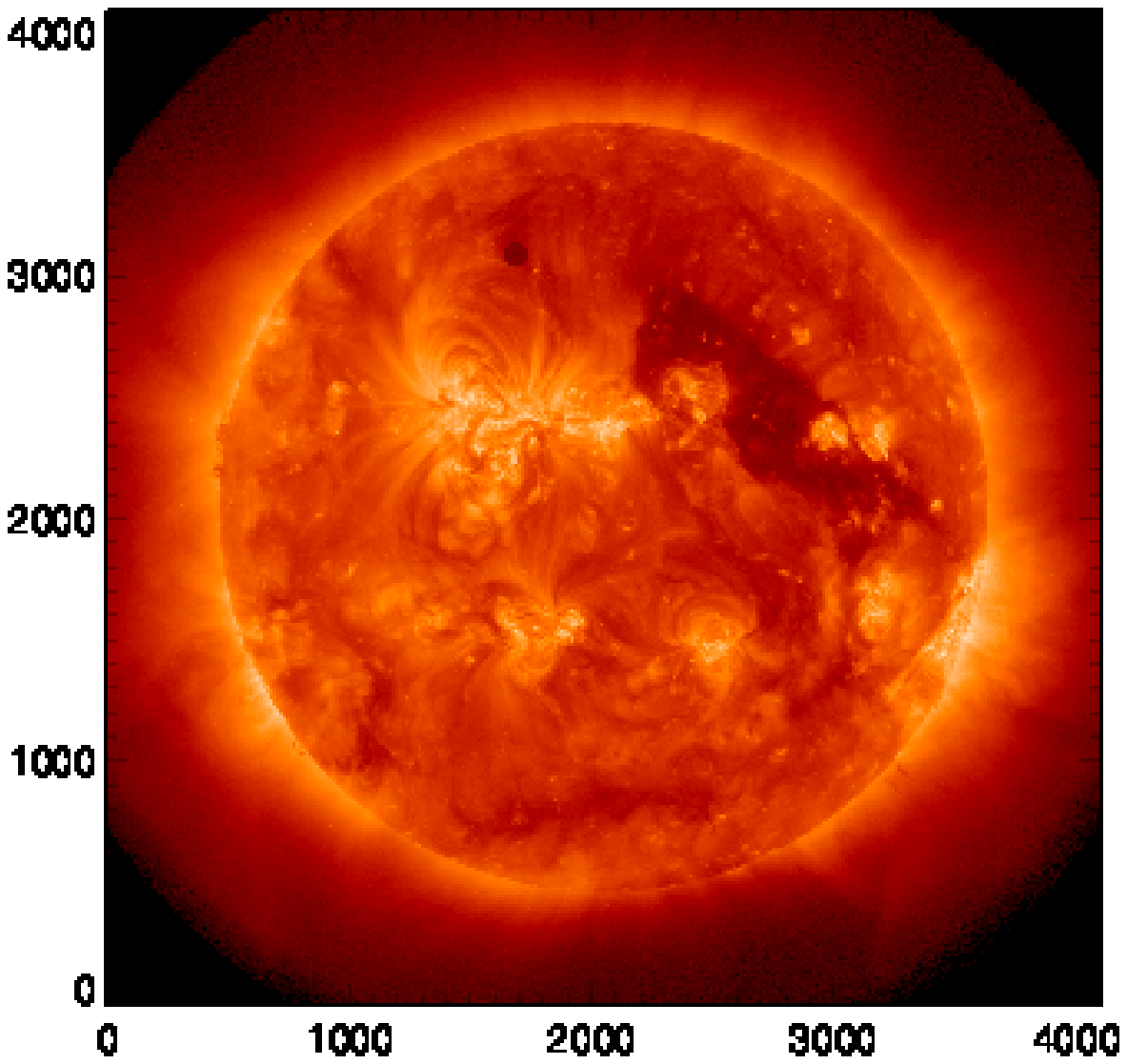} 
\includegraphics[scale=.45]{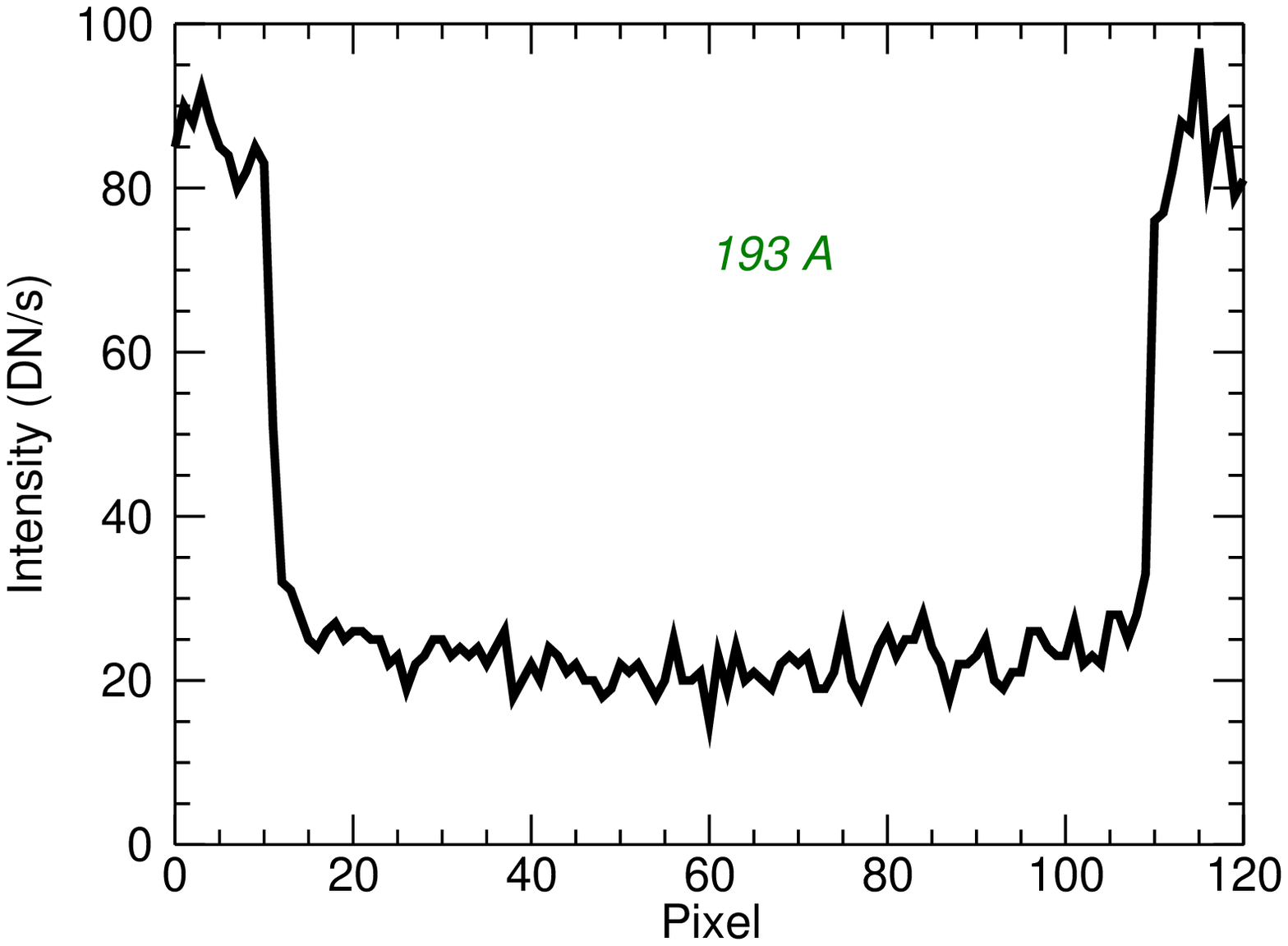} 
 \caption{\small Image in the EUV band (193 \AA) of Venus transit
 (Left) and its horizontal IP (Right).   }
\label{EUVtrans}
\end{figure}

We have done a similar analysis of IPs of Venus\rq\  shadow in the UV and
EUV bands. An image of Venus transit and a sample cross section in the EUV
band, taken with SDO/AIA at {\bf 193} {\AA}, is shown in Fig.~\ref{EUVtrans}.\\

\begin{figure}[!t]         
\includegraphics[scale=1]{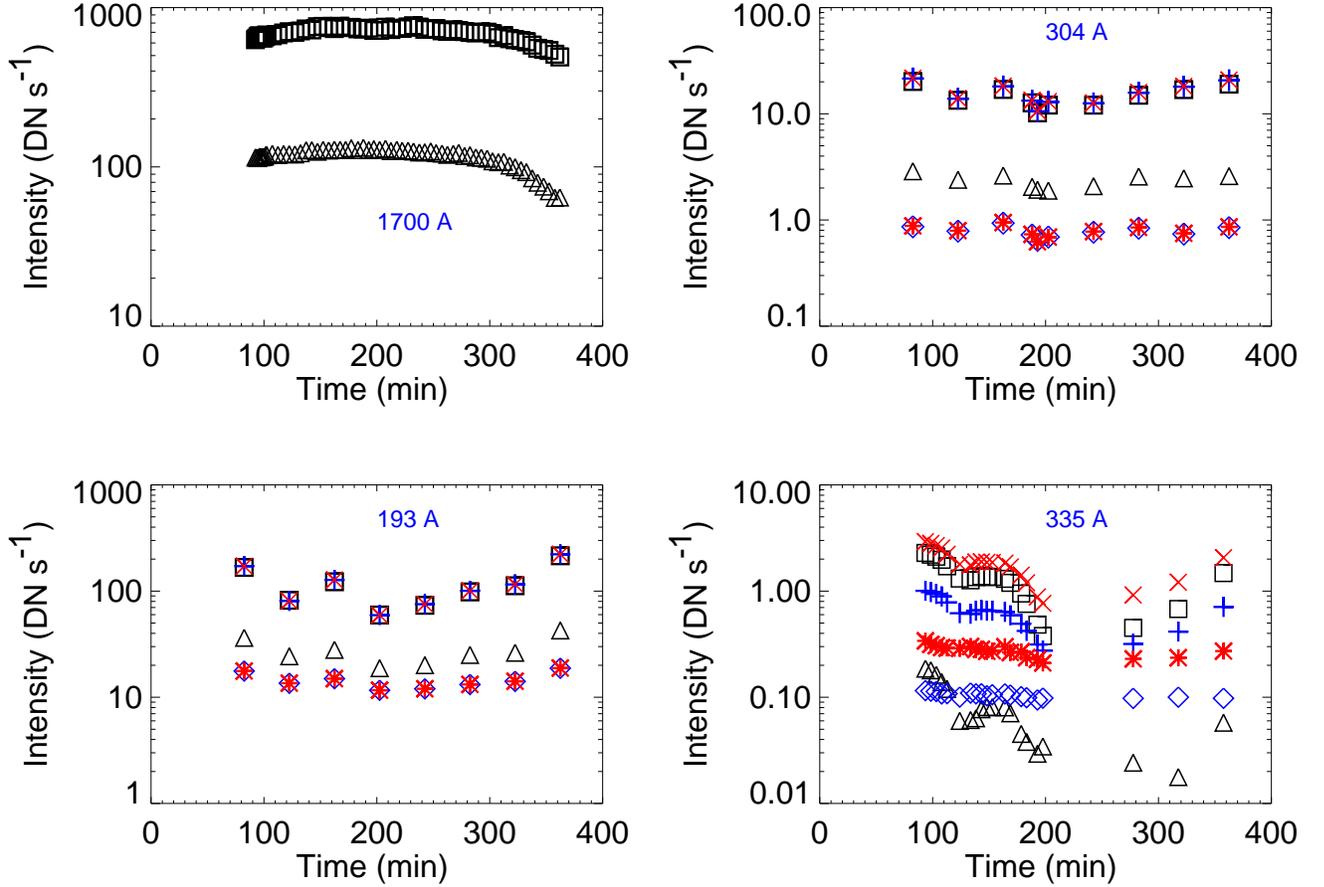}

 \caption{\small Mean flux evolution inside Venus disk and annulus
 1, before and after deconvolution vs. time in UV and EUV bands.
Top Left: 1700 \AA; Top Right: 304 \AA; Bottom Left:193 \AA; Bottom
Right: 335 \AA. Black square: annulus average flux before
deconvolution, red X : annulus average flux after ML deconvolution,
blue cross: same as red X but after aia deconvolution,
black triangle: Venus disk average flux before deconvolution,
red star: Venus disk average flux after M-L deconvolution, blue diamond:
Venus disk average flux after aia deconvolution.}

\label{EUVLC}
\end{figure}

Similarly to what was done for the X-Ray band, we deconvolved the EUV
images to remove the effects of the PSF.  Various works have been
dedicated to deriving the PSF of SDO/AIA. \cite{gri12}
derived the PSF using pre-flight and post-flight measurements and calibrations.
 \cite{Pod13} derived the in-flight SDO/AIA PSF using several
observations, including some of the Moon's limb made during a solar
eclipse observed with SDO/AIA. \cite{Gon16} used observations
of both a solar eclipse and a Venus transit to derive the SDO/AIA
PSF. These authors assumed that there is no emission coming from
the dark side of Venus during the transit, but then discovered that they needed
to include a long range effect, otherwise the parametric PSF
they used would not be able to remove ``the apparent emission inside
the disk of Venus''.\\
Interestingly, in a similar but unrelated study, \cite{Def09} used the
2004 Venus transit to determine the TRACE in-flight PSF. They, too,
assumed that no EUV radiation comes from Venus\rq\ dark side, and then found that
``much more scattered light is found than can be accounted for merely
by diffraction'' and that half of the scattered light was due to some
other mechanism.\\
It is quite possible that both \cite{Gon16} and \cite{Def09} had discovered, and
were trying to account for, some real EUV emission of the kind we find.\\
For the above reasons we decided to adopt the PSF derived in \cite{gri12},
which is available in SSW and is a standard in deconvolving AIA EUV images. We
have also applied the \cite{Pod13} PSF, kindly provided
by the author, to test if results are different.  This latter PSF is not
available for full disk images so we have compared the results obtained
with SSW and Poduval PSF only for partial disk images; we found that we
get in practice the same average flux with the standard deconvolution in
SSW and with the deconvolution which uses Poduval PSF. Being reassured
by this result, we resorted to the \cite{gri12} PSF to deconvolve full disk images.
We concentrated on full disk images (albeit their number is smaller)
because we are thereby certain to remove even possible long range effects.\\
We have, thus, deconvolved each full disk image with the PSF available
in SSW and derived the average flux in Venus shadow and in annulus
1.  In Fig.~\ref{EUVLC} we show the evolution of the UV and EUV
average flux values as observed with SDO/AIA in the dark side of
Venus and in the smallest annulus (annulus 1) taken around Venus,
before and after deconvolution performed with the AIA routine and
with the Maximum Likelihood (M-L) routine.\\
The data in the 335 \AA\ band
show an unusual behaviour with deconvolution: the average flux
inside Venus\rq\ disk increases after deconvolution, just the opposite
of what one expects and which happens in any other band.  The flux
in Venus\rq\ shadow is rather low before deconvolution, so 
any signal that is present and not due to noise must be marginal. Indeed, the signal
after deconvolution is virtually constant. We are thus forced to
consider the results for 335 \AA\ images as unreliable; we present
all the relevant results just for completeness and as an additional
null test, but these results are of little relevance for our problem.\\
No PSF is available for the 1700 \AA\ band,
so we cannot deconvolve the relevant data; we simply used the non-deconvolved
data.

\begin{figure}[!hbt]               
\centering
  \includegraphics[scale=1]{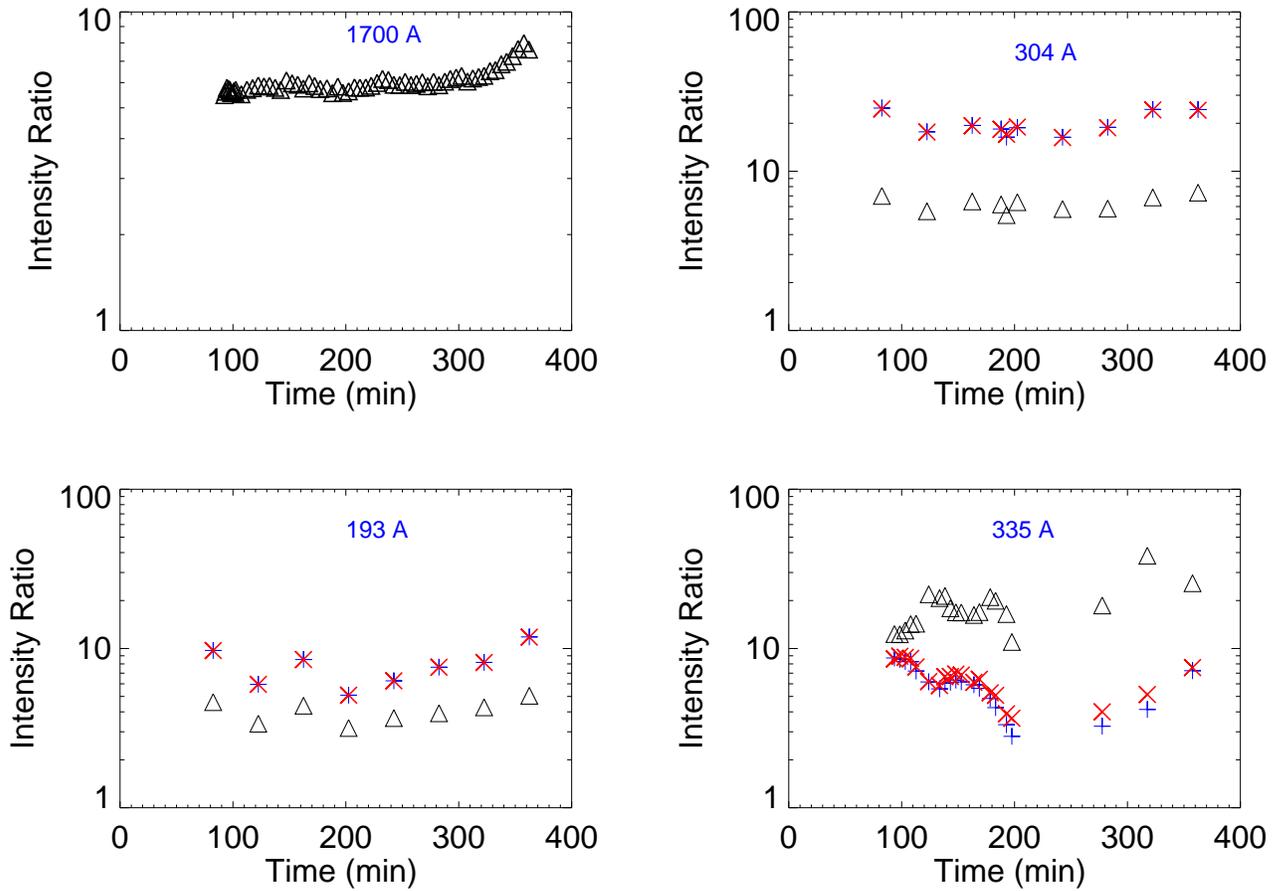}

 \caption{\small Evolution of (mean annulus 1 flux to  mean Venus
 disk flux) ratio before and after deconvolution in UV and EUV bands. Top
 Left: 1700 \AA; Top Right: 304 \AA; Bottom Left:193 \AA; Bottom Right:
 335 \AA. Black Triangle: Ratio before deconvolution; Blue cross: Ratio
 after aia deconvolution; Red X: Ratio after ML deconvolution. }

\label{ratiosevolv}
\end{figure}

\begin{figure}[!hbt]               
\centering
   \subfigure[]{\includegraphics[width=8cm]{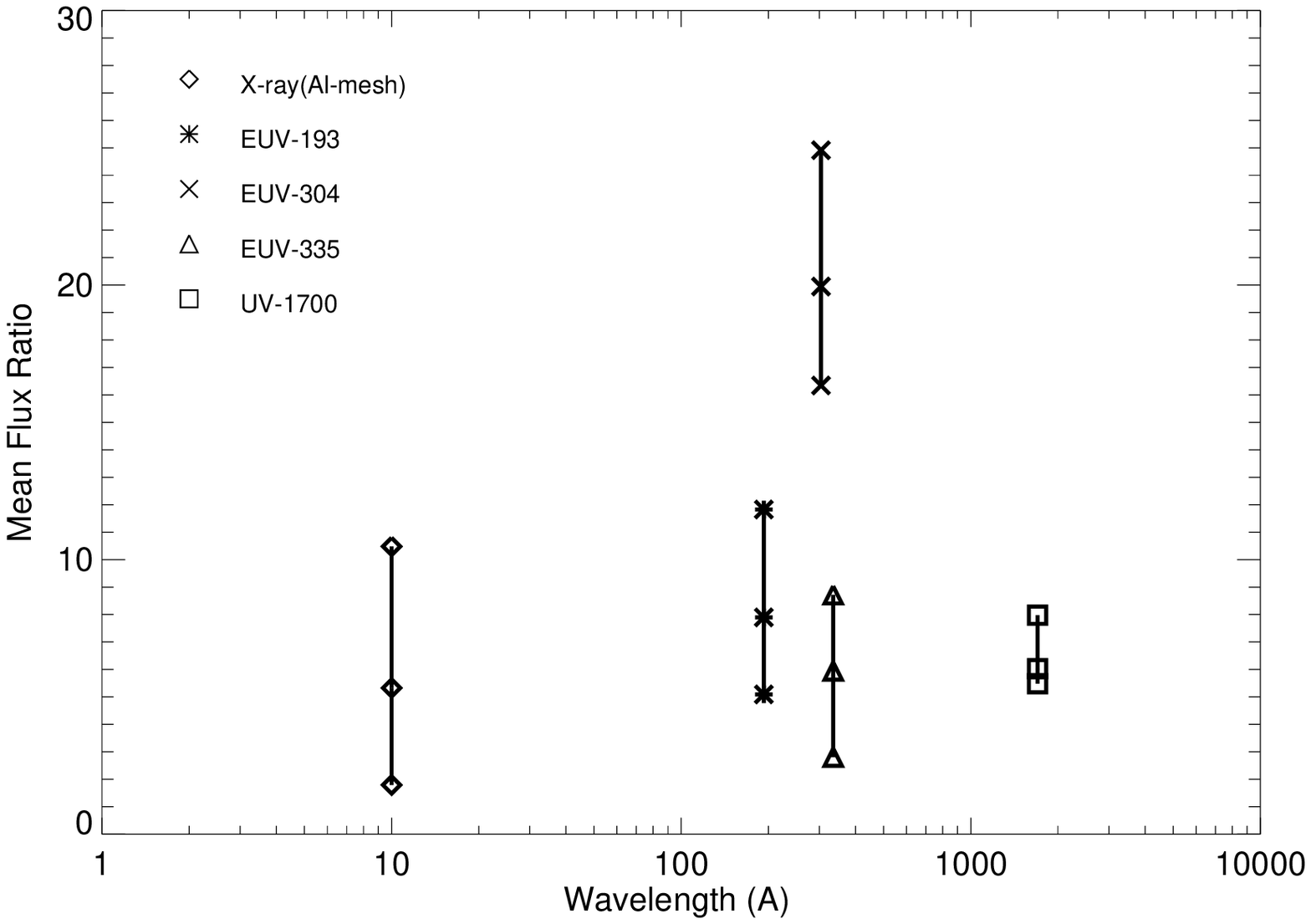}}
 \subfigure[]{\includegraphics[width=8cm]{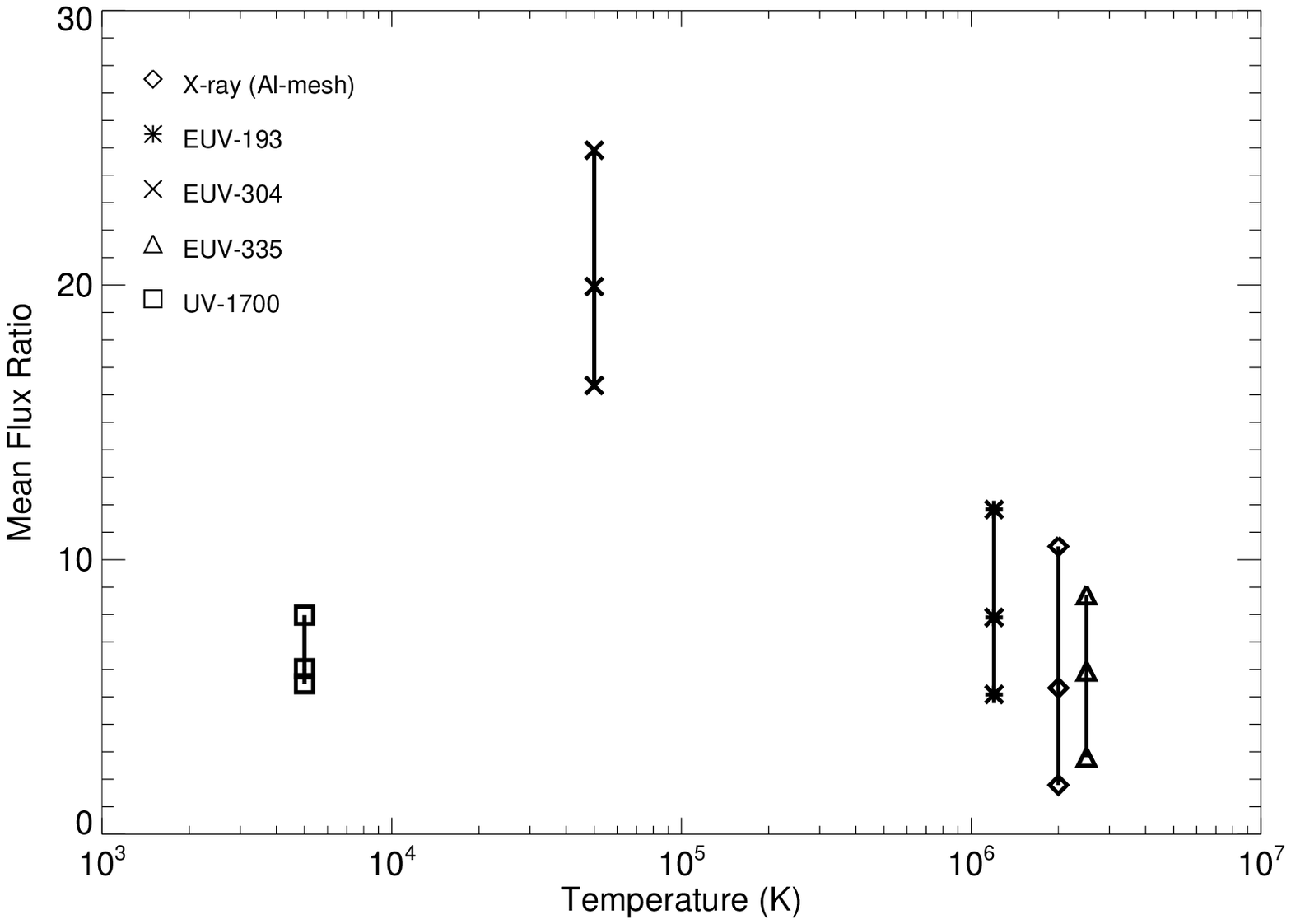}}
 
 \caption{\small Minimum, maximum and mean values of the ratio between the average intensity on solar disk around Venus and
average flux inside Venus\rq\  shadow vs.\ wavelength (left panel) and vs.\ solar plasma
temperature (right panel).}

\label{ratios}
\end{figure}

It is immediately apparent that the flux evolution in any EUV band has no
evident 
correlation with that of the X-Ray flux; in the 1700  {\AA} band the flux
increases as Venus crosses the solar disk, and decreases thereafter. In the
335  {\AA} and 193  {\AA} bands the opposite occurs; few changes are
seen in the 304 {\AA} band. However it appears that also in these cases the
flux in the shadow clearly follows that in the surrounding ring, almost
(but not exactly) by a fixed factor. \\
This approximate proportionality of the flux in Venus\rq\  dark side
relative to the surrounding regions hints at a strong analogy between
the mechanisms generating Venus\rq\  dark side emission in X-ray,  
EUV and UV bands.\\
 Fig.~\ref{ratiosevolv} shows the evolution of the ratio between flux
values inside the annulus and in the Venusian disk, for the original data
and for the images deconvolved with both methods.
\\
Fig.~\ref{ratios} shows the minimum, maximum and mean values of the
ratio between flux values inside the annulus and in the Venus disk
versus wavelength and versus temperature (of the plasma which would
be observed on the Sun). Since any possible deconvolution would slightly
decrease the flux in Venus disk also at 1700 A, we may expect that the
data points at 1700 are higher. There appears to be a slight increasing trend
with wavelength.

\section{Conclusion}  
We studied the Venus transit across the solar disk which occurred in 2012
and was observed with \textit{Hinode}/XRT in the X-Ray band and SDO/AIA in
the EUV and UV bands.  We have measured a significant X-Ray residual
flux from Venus\rq\  dark side (i.e., from the Earth-facing side)
during the transit that was significantly above the estimated noise level
of 2 DN s$ ^{-1} $, as reported by \cite{Kob14}.\\ Let us discuss the
systematic uncertainties of XRT flux.  According to \cite{Kob14} there
are two kinds of systematic uncertainties for XRT. The first are those
which have a reliable quantitative correction procedure such as: dark
current, Fourier, vignetting, and JPEG compression noise sources; their
correction procedures have been successfully embedded in xrt\_prep.pro
(the calibration reformatter). Since all of the data we have used for X-ray
analysis (both Ti-poly and Al-mesh filters) have been prepared with
xrt\_prep.pro, we expect that this class of uncertainties has been properly
corrected and does not explain the observed residual.\\
The second kind of systematic
uncertainties are model-dependent and are not included in xrt\_prep.pro;
among them are light scattering by the grazing-incidence mirror of XRT,
visible straylight leak and photon counting uncertainty \citep{Kob14}.
In this respect the error bars for each flux value have been computed as
follows. From the flux values, the exposure times, and the conversion
factor from DN to photons, we have computed the number of photons
collected and from them the statistical errors due to Poisson noise.
This statistical error has been converted to a flux error bar per
data point.
The resulting error bar is typically less than the 2 DN s$ ^{-1} $ mentioned
by \citep{Kob14}.
In sections 4 and 5 we comprehensively discuss the PSF scattering
and visible light leak effects.\\
To test the performance of the instrument's PSF (i.e., due to
instrumental X-ray scattering) and the possible effect of the atmosphere
on the residual flux, we studied a Mercury transit across the solar disk,
observed with the \textit{Hinode}/XRT in 2006. We measured an apparent
X-Ray residual flux in the case of Mercury before deconvolution.\\
For both Venus and Mercury we used a new version of the
\textit{Hinode}/XRT PSF, selected well illuminated images in the X-Ray band,
and deconvolved them. Even after deconvolution, flux from Venus\rq\  shadow
has remained significant, while in the Mercury case it has become negligible.
So it appears that the observed flux in Venus\rq\  shadow is real.\\
As for the Venus case, we have analyzed two X-ray datasets: a set collected with
Ti-poly filter and another collected with Al-mesh filter. While the
former is potentially strongly affected by a light leak that appeared a short time before
that Venus transit, the latter is not. Both datasets, however, clearly show
the presence of a significant flux from Venus\rq\   dark side, showing the
reality of this effect. Although we consider the results from the
Al-mesh data set to be a strong confirmation for the observed X-Ray residual
flux, the Ti-poly results also provide more confidence about the
observed residual flux and prove this effect in more than one filter. The
level of the residual flux is not constant: as Venus crosses the
solar disk it gradually grows, reaching a maximum roughly halfway through the
transit, and then gradually decreases as it approaches the solar limb. The
flux changes by an almost fixed factor of the flux of the surrounding
solar regions (i.e., along nearby lines of sight) as shown in
Figs.~\ref{LCTiDecon} and \ref{LCDeconAl}.  On the other hand the use of
the PSF and the test on Mercury convincingly shows the removal of any PSF
effect.  Furthermore, any light leak effect would instead be expected to
be almost uniform or constant in time.\\
The PSF of XRT has also been determined at 1.0 keV. 
We find that deconvolving the images with this PSF 
reduces by a factor of about 0.5, on average, the flux inside the Venusian disk. More
specifically,
the mean flux across the disk before deconvolution is about 24
DN s$ ^{-1} $. After deconvolution with the 0.56 keV PSF model it is about 20 DN s$ ^{-1} $, but after deconvolution with the
1.0 keV PSF model it is about 10 DN s$ ^{-1} $.
Therefore a significant flux level still remains,
even after deconvolution with the 1.0 keV PSF, showing the reality of the effect nonetheless.\\
In this respect, however, we believe that the PSF at 0.5 keV is more
appropriate to our study.  In fact, we are detecting photons coming from
the corona and re-processed at Venus or in Venus\rq\  magnetotail (or
something related to Venus), a process which should not
raise photon energy.  The corona is at a few MK (at most 3 or 5, and only
then in some places like active region cores), and no flare
appears during the Venus transit. So a relatively smaller fraction of
photons are expected even at 0.56 keV. We use the Al-mesh and Ti-poly
filters; so considering the coronal spectrum folded with the Al-mesh filter
response (Al-mesh data are the most reliable ones for Venus X-ray
observations), we may shift the average of the observed plasma
emission some 0.1 keVs closer to, but not at, 0.56 keV.  On one
hand we are confident that the general result is robust against the
choice of PSF model, but the use of the 0.56 keV PSF can be
considered to be a conservative evaluation, and so we can use the relevant
results quite safely. \\
The analogous kind of analysis made in four EUV bands observed with
SDO/AIA has shown that there is also some flux in these bands coming
from Venus\rq\ night side, and that its evolution clearly follows that
of the flux inside an annulus surrounding Venus. The light curves
do not show, however, any trend
similar to that of the X-Ray flux.\\
Past X-ray observations of Venus were very different, in many respects.
In January 2001, Venus was observed for the first time with the Chandra X-ray
telescope. \citet{Den02} proposed that the fluorescent scattering of
solar X-rays from Venus\rq\  atmosphere was the primary source of the X-ray
emission they observed. Not only the morphology, but also the observed X-ray
luminosity was consistent with the scattering of solar X-rays \citep{Den02}. \\
In 2006 and 2007 again with Chandra, besides fluorescent scattering, Solar
Wind Charge eXchange (SWCX) emission was clearly detected.
Comparison of X-ray images taken in 2006 and 2007 with those obtained in 2001
(taken at a similar phase angle) showed that the limb brightening had
increased. This would be the case if the X-ray radiation from Venus was the
superposition of scattered solar X-rays and SWCX emission. The lack of
detection of any SWCX-induced X-ray halo in the first Venus observation
was explained by being during a high level of the solar X-ray cycle
\citep{den08}.\\
Previous X-ray observations, however, have shown X-ray emission from the
sunlit side of Venus. The low intensity we detect
in X-ray and EUV comes from the dark side of Venus, and appears to have
a totally different origin; it appears to evolve
during the transit remaining, at any time, approximately proportional to
the emission of the solar regions along nearby lines of sight.
This intensity cannot be due to scattering in the upper atmosphere of
Venus because we should detect a brighter inner rim in Venus\rq\  shadow.\\
The effect we are observing could be due to scattering or re-emission
occurring in the shadow or wake of Venus. One possibility is due to the very long magnetotail of Venus, ablated by the solar
wind and known to reach Earth's orbit \citep{Gru97}.
This magnetotail could be side-illuminated from the surrounding regions
and could scatter, or re-emit, the radiation; the cone of Venus
shadow reaches up to $ 9.6 \times 10^5$ km away from Venus, leaving ample
space ($\approx 4.5 \times 10^7 $ km) for side-illuminating the  magnetotail.
The emission we observe would be the reemitted radiation integrated
along the magnetotail.\\
One wonders if such an effect is important for exoplanets, in particular
for those Jupiter-size planets orbiting very close to their stars; they
may have a very large ablated tail, especially if they do not have a
magnetic field. To some extent, the study of these tails may help to
understand, among other issues,
the presence (or lack thereof) of magnetic fields.\\
Future work will study in more detail this phenomenon: we plan to study
some faint structures present in the shadow and address
possible physical mechanisms involved in generating the residual
emission. \\

\acknowledgments{ \textbf{Acknowledgments} \newline
We thank an anonymous referee for suggestions and comments on EUV deconvolution.
M.A., G.P., A.P., F.R. acknowledge support from Italian Ministero
dell'Universit\`a e Ricerca; P.J. and M.W. were supported under contract
NNM07AB07C from MSFC/NASA to SAO.  Some of the routines for the data
analysis and some early evaluations were kindly supplied by A. F.
Gambino. SDO data were supplied courtesy of the SDO/AIA consortia. SDO
is the first mission to be launched for NASA's Living With a Star
Program. Hinode is a Japanese mission developed and launched by ISAS/JAXA,
with NAOJ as domestic partner and NASA and STFC (UK) as international
partners. It is operated by these agencies in co-operation with ESA and
the NSC (Norway).}

\section*{APPENDIX A}
Metrology data and on-orbit observations are used to model the point spread
function of the X-Ray Telescope's (XRT; \citep{Gol07}) mirror assuming that XRT
is operated at the best on-axis focus. The metrology data estimate encircled
energy profiles for two energies, 0.56 keV and 1.0 keV.  We develop a PSF for
both energies and find the function that returns the encircled energy data as a
piecewise continuous function composed of a Lorentzian core and a series of
power-law functions as its wings. The PSFs we develop do not consider other
sources of scattering such as the effects of changing the focus position, other
elements within the optical system, filters, and the CCD camera system, or
material
contamination on the XRT CCD. We do not incorporate any non-axisymmetric
structures although the system PSF is known to vary (see Fig.~4 in
\cite{Gol07}).\\
The XRT mirror is a Wolter Type-I grazing incident optic built by Goodrich. The
XRT has 9 broadband filters that sample plasma temperatures from 0.5--10 million
Kelvin and is equipped with a 2048x2048 CCD. The XRT has 1.02860 arcsec pixels
with a wide field of view of 34x34 arcmin \citep{Kan08}. The mirror manufacturer
provided encircled energy estimates based on the Power Spectral Density (PSD)
derived from measurements of the mirror surface roughness.\\
We construct two PSFs for the XRT mirror using a semi-empirical approach. We
model the core of the PSF by considering a variety of functions that could
reproduce the on-axis encircled energy data. We then estimate the wings of the
PSF based on scattering patterns observed in XRT data. The wings of the PSF are
modeled assuming a piecewise continuous power law of the form:
$$ P_{i}(r)=r^{-\alpha_{i}}, {\rm where\ } {\alpha_{i}} \geq 0, {\rm and\ } 
{\alpha_{i+1}} \geq {\alpha_{i}}, $$
where $r$ is the radial distance from the optical axis. We assume the PSF is
spatially invariant and only depends on the radial distance from the scattering
source. Scattered light from XRT data are used to determine the breakpoints of
the function so that the following criteria are met:\\

\begin{enumerate}
\item The metrology data affirms that 81\% of the encircled energy lies within
5 arcsec for a 0.56 keV source and 77\% for a 1.00 keV source, meeting design
specifications.
\item The wings of the PSF match and extend the slope of the encircled energy in
a smooth and continuous way.
\item In the case when the PSF will be normalized, we assume that 100\% of the
light will be scattered within the XRT field of view. 
\end{enumerate}

\begin{figure}[!htb]
\centering
\includegraphics[scale=0.5]{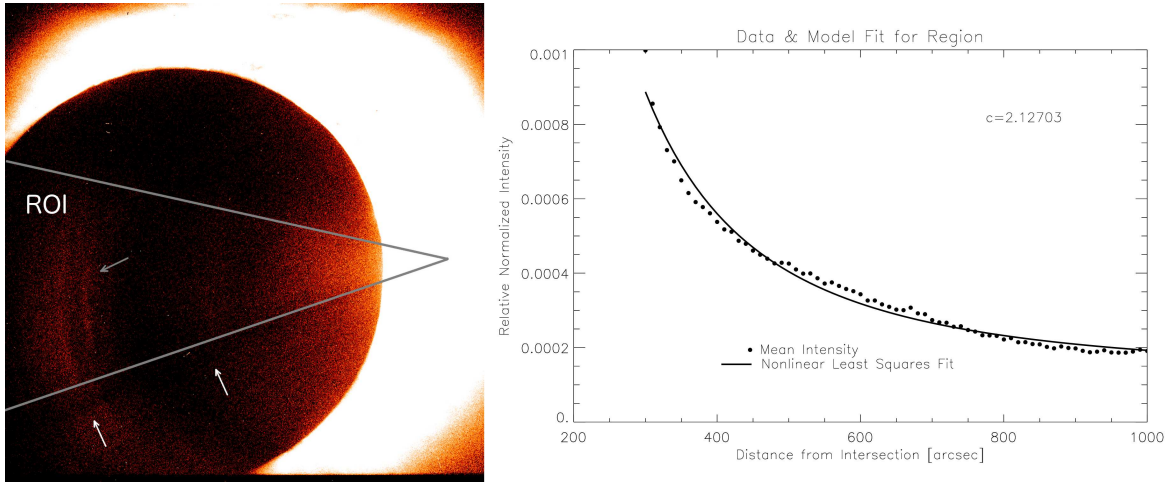}
\caption{\emph{Left}: Hinode XRT C-Poly (log intensity) eclipse image taken 2008 February 7 at 5:17:53 UT. White arrows point to the shadow of the entrance filters and the gray arrow points to the azimuthal scattered light pattern. \emph{Right}: Plot of the mean relative normalized intensity as a function of distance from the intersection of the two lines within the region of interest (ROI).  Within the ROI, the scattered light follows an inverse square law up through the gray arrow when the intensity increases again. This fit does not distinguish between the mirror scatter and the azimuthal scatter.}
\label{Appex_fig1}
\end{figure}

\begin{figure*}[!htb]
\centering
\includegraphics[width=\textwidth]{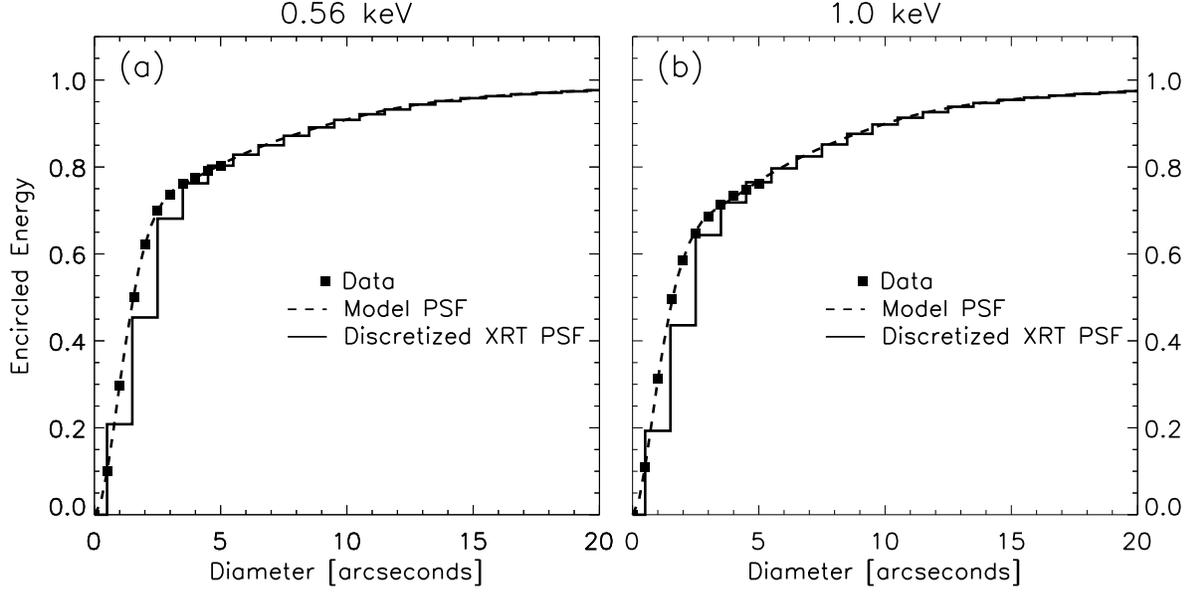}
\caption{Encircled energy plot of the manufacturer data (squares), the model PSF (dashed line) and the model discretized for XRT pixel size (solid line) for 0.56 keV, (a) and 1.0 keV (b).}
\label{Appex_fig2}
\end{figure*}

To gauge how light is scattered far from the source we use full frame images of
(a) limb flare data in which a bright flare occurs on the solar limb when the
Sun is centered in the field of view, and (b) solar eclipse data when the Moon
passes between XRT and the Sun. Because of Hinode's orbit, XRT experiences
either a partial or total solar eclipse twice a year and XRT takes full disk
data in several filters. We use the Moon as a way to measure scattered light
when it partially obscures an active region on the Sun. Scattered light from
these bright regions is easily visible across the Moon's shadow.\\
In addition to the scattered light from the mirror, there are at least two other
causes of scattered light. First is the scattering due to the entrance apertures
that is accentuated during a solar flare, and the second is a pattern of
scattered light that is pointing dependent and is always present. The left panel
of Fig.\ref{Appex_fig1} shows an example of solar eclipse data used in the analysis and
demonstrates both patterns of scattered light within the Moon's shadow. 
The image is scaled on a log scale with pixel values above a few DN s$ ^{-1} $ saturated to
white so that the low-level scatter can easily be seen. Dark bands that appear to emanate from the scattering source are the pattern created from the entrance
filters (white arrows). A gray arrow points to the second scattering pattern. It is a partial dark
ring followed by a region of bright light. They appear as partial bands around
the scattering source. This pattern is pointing dependent and changes location
depending on pointing and the location of the scattering source within the
field of view. \\
All XRT data are processed using the standard reduction routines provided by the
XRT team in SolarSoft. We use full resolution images. The exposure times of the
data vary between 0.5--16 seconds depending on the solar conditions.  To estimate
the amount of scatter in an XRT image, the average normalized intensity along an
arc as a function of radial distance from the scattering source is fit to the
general power law function of the form:
$$ P(r)=\alpha(r-b)^{-c}+d, $$
where $a$, $b$, $c$, and $d$ are all free parameters. The intensity is
normalized to the maximum value set by the data reduction routine, 
xrt$_{-}$prep.pro. This fitting method is not able to distinguish between the
different sources of scatter.\\ 
To mitigate the effects of the scattering due to the entrance filters, we select
a region between dark bands of scattered light.  An example of a region is the
one between the two lines on the image on the left of Fig.\ref{Appex_fig1}. We attempt to
deal with the second source of scattered light by considering regions closer to
the scattering source rather than farther away. \\
The encircled energy data imply the wings of the PSF do not significantly
contribute to the encircled energy far from the center. At a radial distance of
4--5 arcsec there is little increase in the encircled energies. Therefore, we
expect that far from the source, the other scattering elements will dominate the
scattering. \\
We use Mathematica to calculate the encircled energy curves for both PSF models,
corresponding to the two energy channels, over a spatial grid that is
appropriate for the meteorology data and that oversamples the instrument plate
scale. Fig.\ref{Appex_fig2} shows the encircled energy measurements (squares) for 0.56 keV (a) and 1.0 keV (b). 
\newline
For each of the datasets, we find that a single function could not
reproduce the given encircled energies. The two energies have the same
functional form but have different parameter values and breakpoints; the relevant values are given in Table \ref{Tab PSF values}. We find the
inner portion of the PSF is best represented by a Lorentzian function out to an
inner radius, $r_{0}$. From $r_{0} $ to 5 arcsec, the $r^{-1}$ function
returns the correct encircled energy measurements.\\
We then use the assumption that the PSF will continue to follow the power law
trend and fit the following model:\\
\large  
\[
    P(r)= 
\begin{cases}
  
    a \frac{exp(-\frac{r^2}{\sigma^2})}{\gamma^2 \: + \:r^2} ,&  r \leq r_{0} ; \\
    br^{-1},&    r_{0} \leq r  \leq r_{1} ;  \\
    cr^{-2},&    r_{1} \leq r  \leq r_{2} ; \\
    dr^{-4},&    r_{2}  \leq  r \leq \infty .
\end{cases}
\]
\normalsize   

\begin{table}[!htb]
\centering

\caption {Normalized PSF parameter values.\newline
* Denotes exact values and not approximations.} 

\begin{tabular}{ccc} 
\hline
 \textbf{Parameter Values} & \textbf{0.56 keV} & \textbf{1.0 keV} \\
\hline
\rowcolor{gray}
 \textbf{$ \sigma $} & 2.19256 & 2.36982  \\
 \textbf{$  \gamma$} & 1.24891 & 0.914686  \\ 
\rowcolor{gray}
\textbf{  a} & 1.31946  & 0.847955  \\
\textbf{ b} & 0.03* & 0.038*  \\ 
\rowcolor{gray}
\textbf{ c} & 0.15* & 0.19* \\
\textbf{d} & 18.4815* r$ _{0} $ & 20.1571* r$ _{0} $  \\
\rowcolor{gray}
\textbf{r$ _{0} $} & 3.4167 & 3.22857  \\
\textbf{r$ _{1} $ }& 5* & 5*  \\
\rowcolor{gray}
\textbf{r$ _{2} $ }& 11.1* & 10.3*  \\ 
\hline
\end{tabular}
\label{Tab PSF values}
\end{table}
A plot of the encircled energy (squares) for both channels is given in Figure 2 along with the model PSF (black lines). The models fit the data well. We also discretized the models to the XRT pixel size (solid line). \\
A simple calculation is applied to consider the relative applicability of the
two PSF models for a range of typical plasma temperatures in the corona. We make
use of the Astrophysical Plasma Emission Code (APEC, \cite{Smi01}) to model the
plasma emission as a function of wavelength and temperature. We fold this model
through the XRT's spectral response and convert the instrument spectral response
to a temperature response for each of XRT's filters. We create a spectral
response of several plasma temperatures and compare the amount of energy at or
below 0.75 keV to the amount of energy above 0.75 keV for a given temperature
plasma. Table \ref{Tab. XRT response} shows the relative spectral response for each of the XRT's
filters. \\
 \begin{table}[ht]
\centering

\caption {Relative spectral response for each of the XRT's filters assuming the specified temperature.} 
\begin{tabular*}{\textwidth}{c @{\extracolsep{\fill}} cccccccc}

\hline
 & \textbf{1MK} & \textbf{1MK} & \textbf{3MK}& \textbf{3MK} & \textbf{5MK}& \textbf{5MK} & \textbf{10MK}& \textbf{10MK}\\
\hline
\rowcolor{gray}
 \textbf{Filter} & \textbf{$ \leq $ 750ev} & \textbf{$ > $ 750ev} & \textbf{$ \leq $ 750ev} & \textbf{$ > $  750ev} & \textbf{$ \leq $ 750ev} & \textbf{$ > $  750ev}& \textbf{$ \leq $ 750ev} & \textbf{$ > $  750ev}\\
Al-mesh &0.98 &0.02 &0.23 &0.77& 0.07& 0.93& 0.04& 0.96\\
\rowcolor{gray}
Al-poly \:& 0.95& 0.05&0.17 &0.83  &0.05 & 0.95&0.03 & 0.97\\
C-poly& 0.95& 0.05&0.13 &0.87  &0.03 &0.97 &0.01 & 0.99 \\
 \rowcolor{gray}
Ti-poly\:& 0.96&0.04 &0.13 &0.87  & 0.03&0.97 &0.02 &0.98   \\
Be-thin &0.62 &0.38 & 0.02& 0.98 &0.00 &1.00 &0.00 & 1.00 \\
 \rowcolor{gray}
Be-med\: \: &0.08 &0.92 &0.00 & 1.00 &0.00 &1.00 & 0.00& 1.00  \\
Al-med &0.04 &0.96 &0.00 & 1.00 & 0.00 &1.00 & 0.00& 1.00  \\
 \rowcolor{gray}
Al-thick\: \: &0.00&1.00 &0.00  & 1.00 &0.00 &1.00 & 0.00& 1.00  \\
Be-thick & 0.00 &1.00 &0.00 & 1.00 & 0.00 &1.00 & 0.00& 1.00  \\
\hline  
\end{tabular*}
\label{Tab. XRT response}
\end{table}
Table \ref{Tab. XRT response} shows that, for plasma temperatures above 1MK, a significant portion of
the signal will come from energies greater than 0.75 keV.\\
The PSFs provided above are designed with normalization in mind but this
condition is not necessary, and in fact it forces that 100\% of the energy is
scattered within the XRT field of view. With this condition relaxed, the PSF
will scatter light far from the field of view. Table \ref{Tab PSF} provides the PSF models
without normalization. The only difference between these and the normalized
models is the value of $r_{2} $. This causes the slope of the encircled energy
to essentially remain flat beyond 5 arcsec. 

\begin{table}[ht]
\centering
\caption {PSF model without Normalisation.\newline
* Denotes exact values and not approximations.} 
\begin{tabular}{cccccccc}

\hline
 \textbf{Parameter Values} & \textbf{0.56 keV} & \textbf{1.0 keV} \\
\hline
\rowcolor{gray}
 \textbf{$ \sigma $} & 2.19256 & 2.36982  \\
 \textbf{$  \gamma$} & 1.24891 & 0.914686  \\ 
\rowcolor{gray}
\textbf{r$ _{0} $} & 3.4167 & 3.22857  \\
\textbf{  a} & 1.31946  & 0.847955  \\
\rowcolor{gray}
\textbf{r$ _{1} $} & 5* & 5*  \\
\textbf{ b} & 0.03* & 0.038*  \\ 
\rowcolor{gray}
\textbf{ c} & 0.15* & 0.19* \\
\textbf{ D} & 7.35* r$ _{0} $ & 10.1251* r$ _{0} $  \\
\rowcolor{gray}
\textbf{r$ _{2} $} & 7 & 7.3  \\ 
\textbf{EE at edge of FOV} & 93\% & 93\%\\

\hline
\end{tabular}
\label{Tab PSF}
\end{table}

\section*{APPENDIX B} 
Among different indirect methods of deconvolution available in SolarSoft IDL
libraries, we used codes based on the Maximum Likelihood and Richardson-Lucy
methods:

\begin{description}

\item[$\bullet$]
\textbf{AIA\_ DECONVOLVE\_ RICHARDSONLUCY.pro (AIA)}  based on
Richardson-Lucy algorithm.\\
/darts.isas.jaxa.jp/pub/ssw/sdo/aia/idl/psf/PRO/aia\_deconvolve\_richardsonlucy.pro

The Richardson-Lucy algorithm in this code follows closely the algorithm
discussed by \cite{Jan97}.
\item[$\bullet$]\textbf{MAX\_LIKELIHOOD.pro (M-L)} based on Maximum likelihood algorithm.\\
idlastro.gsfc.nasa.gov/ftp/pro/image/max\_likelihood.pro\\
Based on papers by \cite{Ric72} and \cite{Luc74}.

\end{description}



\end{document}